\documentclass[lettersize,journal]{IEEEtran}
\usepackage{amsmath,amsfonts}
\usepackage{algorithmic}
\usepackage{algorithm}
\usepackage{array}
\usepackage[caption=false,font=normalsize,labelfont=sf,textfont=sf]{subfig}
\usepackage{textcomp}
\usepackage{stfloats}
\usepackage{url}
\usepackage{verbatim}
\usepackage{graphicx}
\usepackage{cite}
\usepackage[numbers,sort&compress]{natbib}
\usepackage{rotating}
\usepackage{float}
\usepackage{svg}
\usepackage{amsmath}
\usepackage{amssymb}
\usepackage{xspace}
\usepackage{graphicx}
\usepackage{verbatim}
\usepackage[T1]{fontenc}
\usepackage{xcolor}
\usepackage{soul}
\usepackage{amsmath}
\usepackage{authblk}
\usepackage[hyperfootnotes=false, hidelinks]{hyperref}
\usepackage{pdflscape}
\usepackage{placeins}
\usepackage{adjustbox}
\usepackage{orcidlink}

\usepackage[numbers,sort&compress]{natbib}

\graphicspath{{./Images/}}

\makeatletter
\newcommand*{\rom}[1]{\expandafter\@slowromancap\romannumeral #1@}
\makeatother

\bstctlcite{IEEEexample:BSTcontrol}
\begin{document}
\title{Sequential Experimental Design for X-Ray CT Using Deep Reinforcement Learning\\}

\author{Tianyuan Wang \orcidlink{0009-0001-8245-2363}, Felix Lucka \orcidlink{0000-0002-8763-5177}, and Tristan van Leeuwen \orcidlink{0000-0002-8794-6426} \thanks{These authors are with the Computational Imaging group, Centrum Wiskunde $\&$ Informatica, Amsterdam, The Netherlands (e-mail: tianyuan.wang@cwi.nl; Felix.Lucka@cwi.nl; T.van.Leeuwen@cwi.nl). Tristan van Leeuwen is also with Utrecht University, Mathematical Institute, Utrecht, 3584 CD, The Netherlands.}\thanks{This paper has supplementary downloadable material available at http://ieeexplore.ieee.org., provided by the author. The material includes additional figures and parameter settings for numerical experiments.}}

\markboth{IEEE TRANSACTIONS ON COMPUTATIONAL IMAGING, ~VOL. ~10, 2024}%
{Shell \MakeLowercase{\textit{et al.}}: A Sample Article Using IEEEtran.cls for IEEE Journals}


The article below has been accepted for publication in IEEE Transactions on Computational Imaging. The version in this
pdf includes all revisions by the authors based on peer reviewer suggestions, but it does not include copyediting, proofreading,
and formatting by IEEE. The final published version is available at https://ieeexplore.ieee.org/document/10572344 and has
DOI https://doi.org/10.1109/TCI.2024.3414273.

© 2022 IEEE. Personal use of this material is permitted.
Permission from IEEE must be obtained for all other uses, in
any current or future media, including reprinting/republishing
this material for advertising or promotional purposes, creating
new collective works, for resale or redistribution to servers or
lists, or reuse of any copyrighted component of this work in
other works.

\maketitle

\begin{abstract}
In X-ray Computed Tomography (CT), projections from many angles are acquired and used for 3D reconstruction. To make CT suitable for in-line quality control, reducing the number of angles while maintaining reconstruction quality is necessary. Sparse-angle tomography is a popular approach for obtaining 3D reconstructions from limited data. To optimize its performance, one can adapt scan angles sequentially to select the most informative angles for each scanned object. Mathematically, this corresponds to solving an optimal experimental design (OED) problem. OED problems are high-dimensional, non-convex, bi-level optimization problems that cannot be solved online, i.e., during the scan. To address these challenges, we pose the OED problem as a partially observable Markov decision process in a Bayesian framework, and solve it through deep reinforcement learning. The approach learns efficient non-greedy policies to solve a given class of OED problems through extensive offline training rather than solving a given OED problem directly via numerical optimization. As such, the trained policy can successfully find the most informative scan angles online. We use a policy training method based on the Actor-Critic approach and evaluate its performance on 2D tomography with synthetic data.

\end{abstract}


\section{Introduction}

\IEEEPARstart{X}{-ray} Computed Tomography (CT) is a non-destructive method widely used to evaluate the quality of complex internal structures in industrial parts. However, there is a trade-off between high-quality reconstruction and scanning speed, as a time-consuming full 360-degree rotation is typically needed to obtain comprehensive information. Kazantsev \cite{kazantsev1991information} and Varga \textit{et al.} \cite{varga2011projection} have pointed out that angles are not equally informative. Therefore, reducing the number of angles by extracting more informative data can help to improve the trade-off between reconstruction quality and scanning efficiency. This trade-off can be formulated as a bi-level optimization problem with respect to angle parameters \cite{ruthotto2018optimal}. The low-level optimization problem formulates the image reconstruction based on the chosen, limited projection data, while the high-level optimization problem finds angles that optimize the reconstruction quality.

Bayesian Optimal Experimental Design (OED) is a mathematical framework that enables the acquisition of informative measurements while minimizing experimental costs \cite{lindley1972bayesian, ryan2016fully}. In Bayesian OED, the prior distribution represents the current belief about the underlying ground truth, while the posterior distribution refers to the updated belief after taking into account the new measurements obtained through the selected design. The difference between the prior and updated posterior reflects the change in uncertainty or equivalently the amount of information gained from the experiments. 
In \emph{simultaneous} experimental design, we apply this procedure to select the optimal viewing angles in a single step, while in \emph{sequential} experimental design, the goal is to select the viewing angles step-by-step, based on the projection data that has been collected so far \cite{yin2021end}. It is this variant of the experimental design problem that we are interested in, as it can adapt the selected viewing angles to the object under investigation.

Two widely used methods for measuring the uncertainty reduction or information gain in Bayesian OED are D-optimality, and A-optimality \cite{burger2021sequentially}. D-optimality measures the information gain using the Kullback-Leibler divergence to compare the posterior and prior distributions, while A-optimality computes the expected error between the underlying ground truth and the reconstruction. 

However, the high dimensionality, computational cost, and typically unknown or unobtainable prior distribution prevents the direct application of the aforementioned technique for sequential optimal design in real-time CT imaging. 

Several methods have been proposed to address these issues. Implicit prior information has been the focus of some researchers. To this end, Batenburg \textit{et al.} \cite{batenburg2013dynamic} and Dabravolski \textit{et al.} \cite{dabravolski2014dynamic} used a set of template images comprising Gaussian blobs to represent prior distribution samples and introduced an upper bound \cite{batenburg2011bounds} to approximate the information gain, indicating the solution set's diameter. Gaussian distribution has been used as a tractable method for the prior distribution in \cite{burger2021sequentially}. Burger \textit{et al.} sequentially selected the projection angle and the source-receiver pair's lateral position considering a specific region of interest and explored Bayesian A- and D-optimality to update the posterior in the covariance matrix and mean after each experimental step. Helin \textit{et al.} \cite{helin2022edge} extended this work to non-Gaussian distributions and employed a Total Variation (TV) prior to enhance edges. In practice, a lagged diffusivity iteration generated a series of Gaussian approximations for the TV prior. Additionally, Barbano \textit{et al.} \cite{barbano2022bayesian} proposed a linearized deep image prior that incorporated information from the pilot measurements as a data-dependent prior. They then used a conjugate Gaussian-linear model to determine the next informative angles sequentially. However, these methods can be time-consuming and are not well-suited for fast in-line applications.

In an industrial context, the use of Computer-Aided Design (CAD) models is a common form of prior information. CAD models enable offline optimization by allowing angle acquisition using simulation tools. Fischer \textit{et al.} \cite{fischer2016object} used a CAD model of the object to optimize task-specific trajectories based on the detectability index proposed by Stayman \textit{et al.} \cite{stayman2013task}. The detectability index is computed using the modulation transfer function and noise power spectrum to evaluate its fitness with a user-defined frequency template. In addition to task-specific optimization, Herl \textit{et al.} \cite{herl2021task} considered data completeness optimization using a Tuy-based metric. Meanwhile, Bussy \textit{et al.} \cite{bussy2023fast} obtained a complete set of angles using either a simulation model or a CAD model and then used the discrete empirical interpolation method and related variants to sub-sample from the set of angles. Once a trajectory is optimized offline sequentially by a CAD model, it can be applied fast in the real application. Nonetheless, the alignment of the optimized trajectory outcome to the real-world coordinate system through proper registration is crucial before executing the real scan \cite{bauer2022practical}. Hence, these methods lack genuine adaptability in in-line applications.

In terms of the methods discussed above, achieving adaptivity while maintaining a fast scan for in-line settings still presents a significant challenge. Additionally, informative angles are typically selected in a greedy manner after evaluating all available angle candidates. In the field of medical CT, Shen \textit{et al.} \cite{shen2022learning} addressed this issue by training a deep reinforcement learning agent on a medical CT image dataset to personalize the scanning strategy sequentially. They utilized a gated recurrent unit as a policy network that maps all the previous measurements to a probability distribution over discrete angles and a radiation dose fraction. The next angle is chosen by sampling from this distribution. This way, around $60$ are chosen sequentially. 

\begin{figure}[t]
  \centering
  \includegraphics[width=\linewidth]{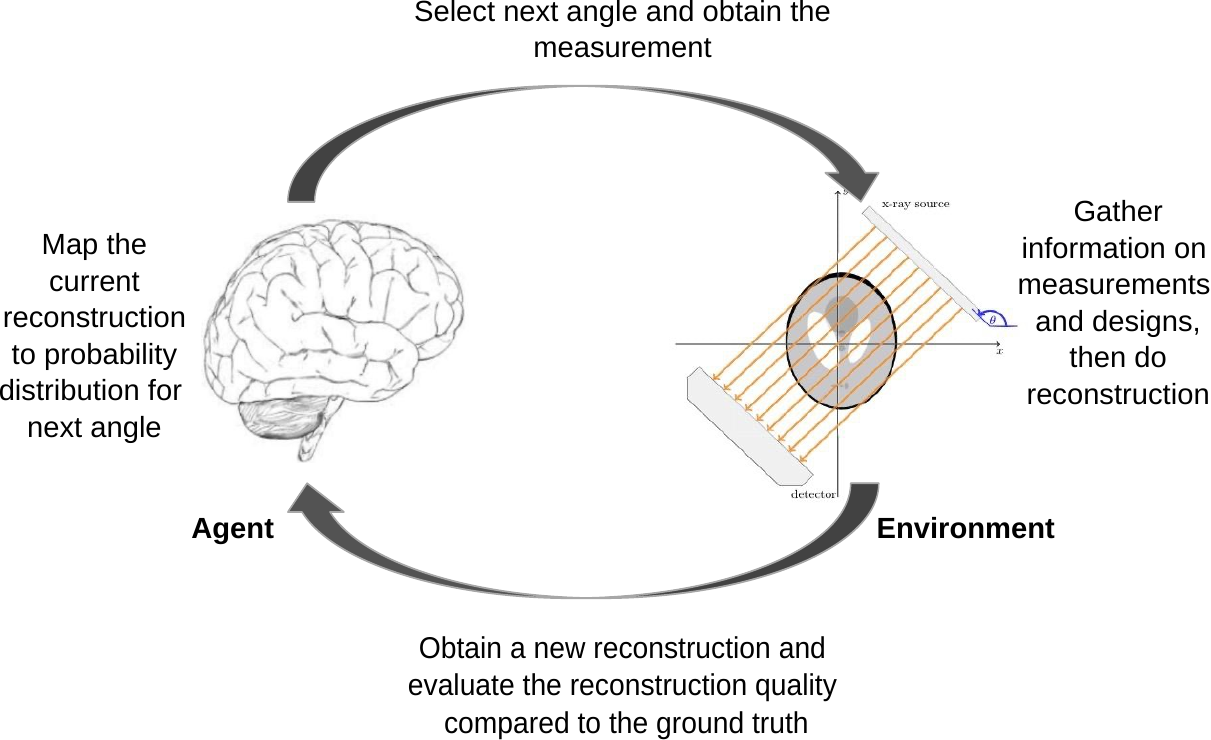}
  \caption{The interaction between the environment and the agent during policy training}
  \label{fig:RL_concept}
\end{figure}

We also leverage deep reinforcement learning to address the aforementioned challenges in our work but we focus on the application of industrial, in-line CT inspection instead of medical CT: We are considering very few scan angles ($<$ 10), simple image features, but a potentially large inter-subject variation due to arbitrary placement and changing samples. For these reasons, we diverge from \cite{shen2022learning} by using the reconstruction space as the main state variable, avoiding problems caused by the increasing number of measurements. Due to this we use very different network architectures to parameterize the learned policy. By employing a deep reinforcement learning approach, we can train the policy to facilitate adaptive angle selection, offering a more efficient alternative to solving the high-dimensional, non-convex, bi-level optimization problem. Figure (\ref{fig:RL_concept}) illustrates the proposed reinforcement learning approach for X-ray CT to solve this OED problem.

The contributions of this work include a novel formulation of the angle selection problem as a Partially Observable Markov Decision Process (POMDP), the use of the Actor-Critic approach from the field of reinforcement learning to address the OED problem, and the development of an adaptive approach that can be fast applied in in-line CT applications.

The structure of this paper is as follows. In section \rom{2}, we present the background on CT reconstruction, Bayesian OED, and reinforcement learning. In section \rom{3}, we discuss the formulation of this experimental design as a POMDP and describe the computation of the policy gradient using the Actor-Critic approach. We provide a set of numerical experiments in section \rom{4} to assess the performance of our proposed method. Finally, in section \rom{5} and section \rom{6}, we discuss and summarize our findings.

\section{Background}

\subsection{CT Reconstruction}
In sparse-angle tomography, the challenge lies in accurately reconstructing an image from incomplete measurement data, where only a limited number of angles are acquired. This inverse problem is severely ill-posed, meaning that small errors in the measurements could result in a large reconstruction error, or that several reconstructions are consistent with the measurements \cite{mueller2012linear}. The Filtered Back-Projection (FBP) algorithm, a traditional analytical reconstruction method, has limitations when used for sparse-angle tomography. It assumes that the measurements are acquired over the full angular range, resulting in inferior reconstructions when applied to limited data \cite{hansen2021computed}.

To address this challenge, it is necessary to incorporate prior information into the reconstruction algorithm to compensate for the limited data \cite{buzug2011computed, hansen2021computed}. Regularised algebraic reconstruction methods have been proposed to incorporate such prior information efficiently. When applied to limited data, these can result in more stable and accurate reconstructions.

Therefore, we represent the object that we would like to reconstruct as $\Bar{\boldsymbol{x}} \in \mathbb{R}^{n}$ where $n \in \mathbb{N}$ represents the number of pixels or voxels. A single noisy measurement $\boldsymbol{y}$ at angle $\theta$ is generated as
\begin{equation}
\boldsymbol{y}(\theta) = \boldsymbol{A}(\theta)\Bar{\boldsymbol{x}} + \boldsymbol{\epsilon},
\label{eq:data model}
\end{equation}
with $\boldsymbol{\epsilon} \sim \mathcal{N}(0, \sigma^2 \boldsymbol{I})$ and $\boldsymbol{A}(\theta)$ is a discretization of the Radon transform along angle $\theta$.

The reconstructed image from $M$ measurements along angles $\boldsymbol{\theta} = \left\{\theta_{1}, \dots, \theta_{M} \right\}$ is obtained via
\begin{equation}
    \widehat{\boldsymbol{x}}(\boldsymbol{\theta}) = \arg \min_{\boldsymbol{x}} {\textstyle{\frac{1}{2}}}\sum_{k=1}^M \left\|\boldsymbol{A}(\theta_k)\boldsymbol{x}-\boldsymbol{y}(\theta_k)\right\|_{2}^{2} + \alpha L(\boldsymbol{x}),   
    \label{eq:reconstruction}
\end{equation}
where $L(\boldsymbol{x})$ is a regularization term representing prior information for $\boldsymbol{x}$.

\subsection{Bayesian OED}
Bayesian OED is a statistical framework that optimizes the design of an experiment by trading off the information gain with the cost of an experiment.

In the context of X-ray CT experimental design, the utility function in Bayesian OED measures the reconstruction quality, where the true underlying ground truth $\Bar{\boldsymbol{x}}$ is estimated by $\widehat{\boldsymbol{x}}(\boldsymbol{\theta})$ from measurements $\boldsymbol{y} \in \mathcal{Y}$ obtained under experimental conditions specified by $\boldsymbol{\theta} \in \mathcal{D}$. The optimal design $\boldsymbol{\theta}^{*}$ maximizes the expectation of the utility function over the design space $\mathcal{D}$ with respect to the measured data $\boldsymbol{y}$ and the model parameter $\Bar{\boldsymbol{x}}$.

Sequential OED is an approach that adjusts the design parameters as new data is acquired. This is achieved by treating the experiment as a sequential decision-making process, where the aim is to select the most informative design parameters based on the observed data to maximize the utility function.  In the $k_{th}$ step of an X-ray CT experiment, the process involves generating observed data using a data model $\pi_{\text{data}}(\boldsymbol{y}_{k}|\Bar{\boldsymbol{x}};\theta_{k})$ (as shown in Equation (\ref{eq:data model})), updating the posterior distribution of $\boldsymbol{x}$ given the observed data up to step $k$ (denoted by $\pi_{\text{post}}(\boldsymbol{x}_{k}|\boldsymbol{y}_{1:k}; \boldsymbol{\theta}_{1:k})$ in Equation (\ref{eq:reconstruction})), obtaining the reconstruction for the underlying ground truth.  Subsequently, the most informative angle $\theta_{k+1}$ is selected as the next design parameter to be used, which maximizes the utility function.

\subsection{Reinforcement Learning}

Reinforcement learning is a widely used approach for sequential decision-making, allowing agents to learn how to map the current state to actions that maximize the total reward for the entire process \cite{sutton2018reinforcement}. Since it considers the long-term effects of actions, reinforcement learning can realize non-greedy sequential decision-making. This approach is based on the Markov Decision Processes (MDPs) framework $\left\{\mathcal{S}, \mathcal{A},  \pi_{t}, R \right\}$, which consists of a set of states $\mathcal{S}$, a set of actions $\mathcal{A}$, a transition operator $\pi_{t}$ representing the conditional probability distribution from the current state to the next state after selecting an action, and a reward function: $\mathcal{S} \times \mathcal{A} \rightarrow \mathbb{R}$ that provides feedback from the environment at each time step. 

A policy $\pi_{\text{policy}}$ in reinforcement learning is a mapping from the current state to a probability distribution of actions: $\pi_{\text{policy}}(a_{k}|\boldsymbol{s}_{k})$. 

In MDPs with a finite number of states, the process begins from an initial state $\boldsymbol{s}_{1}$ with a probability distribution $\pi_{s}(\boldsymbol{s}_{1})$. The agent follows a policy that maps the initial state to the first action, leading the agent to transition to the next state and receive a reward from the environment. This process is repeated until a terminal state is reached, generating a trajectory or an episode $\boldsymbol{\tau} = (\boldsymbol{s}_{1}, a_{1}, r_{1},...,\boldsymbol{s}_{M}, a_{M}, r_{M})$ of $M$ steps.

A Partially Observable Markov Decision Process (POMDP) is an extension of a MDP. In many practical scenarios, an agent may not have full visibility or knowledge about the environment's state. POMDPs come into play in such situations, allowing an agent to make decisions based on limited or partial observations. A POMDP can be defined as a tuple $\left\{\mathcal{S}, \mathcal{A}, \mathcal{O}, \pi_{t}, \pi_{e}, R\right\}$, where two additional components are included in addition to the ones in the standard MDP formulation: a finite observation set $\mathcal{O}$ and an observation function $\pi_{e}$ that defines the conditional probability distribution over the observation in the underlying state after executing an action. Since the agent has limited knowledge about the underlying state in POMDPs, the policy must either map historical observations to the next action or extract information from historical observations in the form of a belief state.

Reinforcement learning aims to find the optimal policy with parameters $\boldsymbol{w}$, denoted as $\pi^{*}_{\text{policy}}(.;\boldsymbol{w})$, that generates the trajectory or episode $\boldsymbol{\tau}$ to maximize the expected total reward. The objective function for reinforcement learning can be expressed as follows:

\begin{equation}
\begin{aligned}
&J(\boldsymbol{w}) = \mathbb{E}_{\boldsymbol{\tau} \sim \pi_{\text{chain}}}\sum\limits_{k=1}^{M} \gamma^{k-1} r_{k}, \\
&\text{where\ } \pi_{\text{chain}} = \pi_{s}(\boldsymbol{s}_{1}) \prod\limits_{k=1}^{M}\pi_{\text{policy}}(a_{k}|\boldsymbol{s}_{k};\boldsymbol{w})\pi_{t}(\boldsymbol{s}_{k+1}|\boldsymbol{s}_{k}, a_{k}).
\end{aligned}
\label{eq:RL objective}
\end{equation}

The objective function measures the expected total reward with a discount factor $\gamma \in (0,1]$ to account for future uncertainty, and $\pi_{\text{chain}}$ represents the trajectory generation process by the policy.

The total rewards for one trajectory are obtained after the agent completes an episode. The expectation over all trajectories can be estimated by sampling many trajectories. To enhance the process's efficiency, some reinforcement learning approaches utilize value functions that evaluate the expected future benefits from the $k_{th}$ step following the policy. The state-value function $V(\boldsymbol{s}_{k})$ quantifies the expected cumulative reward from state $\boldsymbol{s}_{k}$, taking into account all possible trajectories following the current policy that start from this state.

\section{Methods}
\subsection{Sequential OED as a POMDP}
We take the reconstruction as a belief state rather than considering measurements as the state, as done in \cite{shen2022learning}. To formulate the problem, we adopt a Bayesian OED framework and model it as a POMDP. The POMDP formulation for the X-ray CT experiment is defined as follows:
\begin{itemize}
    \item Observation space $\mathcal{O}$: The observation space is defined as the set of measurements generated by the data model expressed in Equation (\ref{eq:data model}).
    \item State space $\mathcal{S}$: The ground truth $\Bar{\boldsymbol{x}}$ represents the underlying state. The current reconstruction (belief state) of the underlying state, denoted by $\widehat{\boldsymbol{x}}(\boldsymbol{\theta}_{1:k})$, is obtained using the SIRT algorithm with box constraints \cite{elfving2012semiconvergence} as specified in Equation (\ref{eq:reconstruction}). For ease of notation, we use $\widehat{\boldsymbol{x}}_{k}$ to represent the reconstruction at the $k_{th}$ step.  In addition, we maintain a vector $\boldsymbol{b}_{k}$ to keep track of the angles that have been selected before the $k_{th}$ experiment to prevent repeating the same angles. 
    \item Action space $\mathcal{A}$: The action space is a discrete design space consisting of 180 integer angles from the range $[0^{\circ}, 180^{\circ})$.
    \item Transition function $\pi_{t}$ and observation function $\pi_{e}$: The transition function $\pi_{t}$ is deterministic, as the underlying state remains unchanged. On the other hand, the data model $\pi_{e}$ given by Equation (\ref{eq:data model}) serves as the observation function, from which we only consider measurement samples.
    \item Reward function $R$: The reward function is defined based on the PSNR value between the reconstruction obtained after selecting the angle $\theta_{k}$ and its ground truth. Two reward settings are considered, both of which correspond to A-optimality in Bayesian OED.:
  
    \begin{itemize}
    \item End-to-end setting: The reward is given as follows:
    $$ R(\widehat{\boldsymbol{x}}_{k+1}, \Bar{\boldsymbol{x}})=
    \begin{cases}
         PSNR(\widehat{\boldsymbol{x}}_{k+1}, \Bar{\boldsymbol{x}}) & \text{if $k = M$}\\
         0 & \text{otherwise} 
    \end{cases}$$
    If the fixed number of angles $M$ is reached, the episode terminates, and the final PSNR value is given. Otherwise, the agent receives a reward of $0$.

    \item Incremental setting: The reward is given as follows:
    $$ R(\widehat{\boldsymbol{x}}_{k+1}, \widehat{\boldsymbol{x}}_{k}, \Bar{\boldsymbol{x}})=
    PSNR(\widehat{\boldsymbol{x}}_{k+1}, \Bar{\boldsymbol{x}}) - PSNR(\widehat{\boldsymbol{x}}_{k}, \Bar{\boldsymbol{x}}) $$
    The reward represents the improvement in the current reconstruction quality compared to the previous step.
\end{itemize}
\end{itemize}

\subsection{Actor-Critic method for policy optimization}    
The Actor-Critic method is a novel category in the field of reinforcement learning for computing the policy gradient on the objective function described in Equation (\ref{eq:RL objective}). The proposed approach leverages the concept of value functions and utilizes a state-value function to obtain the expected future rewards at the current state, thereby expediting the learning process. Additionally, this method parameterizes the value function.  The Temporal-Difference (TD) error \cite{sutton2018reinforcement} is employed in this approach, which calculates the discrepancy between the estimated value function for the current state and the sum of the current reward and the discounted estimated value function for the next state. This enables the state-value function to be updated through bootstrapping and provides a direction for policy gradient.

At the beginning of each episode, a zero matrix and a zero vector are used as the initial state and action vector, respectively. The complete algorithm is presented in Algorithm (\ref{alg:alg1}).

\begin{algorithm}[t]
\caption{Actor-Critic.}\label{alg:alg1}
\begin{algorithmic}[1]
\STATE Initialize the policy parameters $\boldsymbol{w}_{1}$ and the value function parameters $\boldsymbol{w}_{2}$ randomly. Set step sizes $ \alpha^{\boldsymbol{w}_{1}} > 0$ 
and $\alpha^{\boldsymbol{w}_{2}} > 0$
\STATE {\textbf{for each episode do:}}
\STATE \hspace{0.5cm} Get a phantom sample $\Bar{\boldsymbol{x}}$ then initialize $\widehat{\boldsymbol{x}}_{1} = \mathbf{0}$ and  \\
\hspace{0.5cm} $\boldsymbol{b}_{1} = \mathbf{0}$ (first state of this episode). 
\STATE \hspace{0.5cm} \textbf{for} $k
 = 1, ..., M$: 
\STATE \hspace{1cm} Select the angle based on the soft-max policy,\\
\hspace{1cm} which maps the inputs to a probability distribution \\
\hspace{1cm} that sums to 1: 
$\theta_{k} \sim \pi_{\text{policy}} (\cdot|\widehat{\boldsymbol{x}}_{k}, \boldsymbol{b}_{k};\boldsymbol{w}_{1})$
\STATE \hspace{1cm} Get new measurements $\mathbf{y}_{k}$ from Equation (\ref{eq:data model}) \\
 \STATE \hspace{1cm} Reconstruct new image $\widehat{\boldsymbol{x}}_{k+1}$ from Equation (\ref{eq:reconstruction}) \\
 \hspace{1cm} and get a new vector $\boldsymbol{b}_{k+1}$ \\
 \STATE \hspace{1cm} Get reward $r_{k}$ using end-to-end setting\\ 
 \hspace{1cm}  $R(\widehat{\boldsymbol{x}}_{k+1}, \Bar{\boldsymbol{x}})$ or incremental 
  setting $R(\widehat{\boldsymbol{x}}_{k+1}, \widehat{\boldsymbol{x}}_{k}, \Bar{\boldsymbol{x}})$ \\
  \hspace{1cm} Estimate the state-values $ \widehat{V}(\widehat{\boldsymbol{x}}_{k}, \boldsymbol{b}_{k};\boldsymbol{w}_{2})$ and \\
 \hspace{1cm} $ \widehat{V}(\widehat{\boldsymbol{x}}_{k+1}, \boldsymbol{b}_{k+1};\boldsymbol{w}_{2})$ using a neural network
\STATE \hspace{1cm} Compute TD error:\\
 \hspace{1cm} $ \delta_{k} \gets r_{k}+ \gamma \widehat{V}(\widehat{\boldsymbol{x}}_{k+1}, \boldsymbol{b}_{k+1};\boldsymbol{w}_{2}) - \widehat{V}(\widehat{\boldsymbol{x}}_{k}, \boldsymbol{b}_{k};\boldsymbol{w}_{2})$ \\
\hspace{1cm} (If $k = M$, $\widehat{V}(\widehat{\boldsymbol{x}}_{k+1}, \boldsymbol{b}_{k+1};\boldsymbol{w}_{2}) = 0$)
\STATE \hspace{1cm} Update policy function parameters $\boldsymbol{w}_{1}$:\\
\hspace{1cm} $\boldsymbol{w}_{1} \gets \boldsymbol{w}_{1} + \alpha^{\boldsymbol{w}_{1}}\nabla_{\boldsymbol{w}_{1}}\log \pi_{\text{policy}}(\theta_{k}|\widehat{\boldsymbol{x}}_{k}, \boldsymbol{b}_{k};\boldsymbol{w}_{1})\delta_{k} $
\STATE \hspace{1cm} Update value function parameters $\boldsymbol{w}_{2}$:\\
 \hspace{1cm} $\boldsymbol{w}_{2} \gets \boldsymbol{w}_{2} + \alpha^{\boldsymbol{w}_{2}}\nabla_{\boldsymbol{w}_{2}}\widehat{V}(\widehat{\boldsymbol{x}}_{k}, \boldsymbol{b}_{k};\boldsymbol{w}_{2})\delta_{k}$
\STATE \hspace{0.5cm} \textbf{end for}
\STATE \textbf{end for}
\end{algorithmic}
\label{alg1}
\end{algorithm}

\subsection{Network architecture}
\begin{figure}[t]
  \centering
  \includegraphics[width=\linewidth]{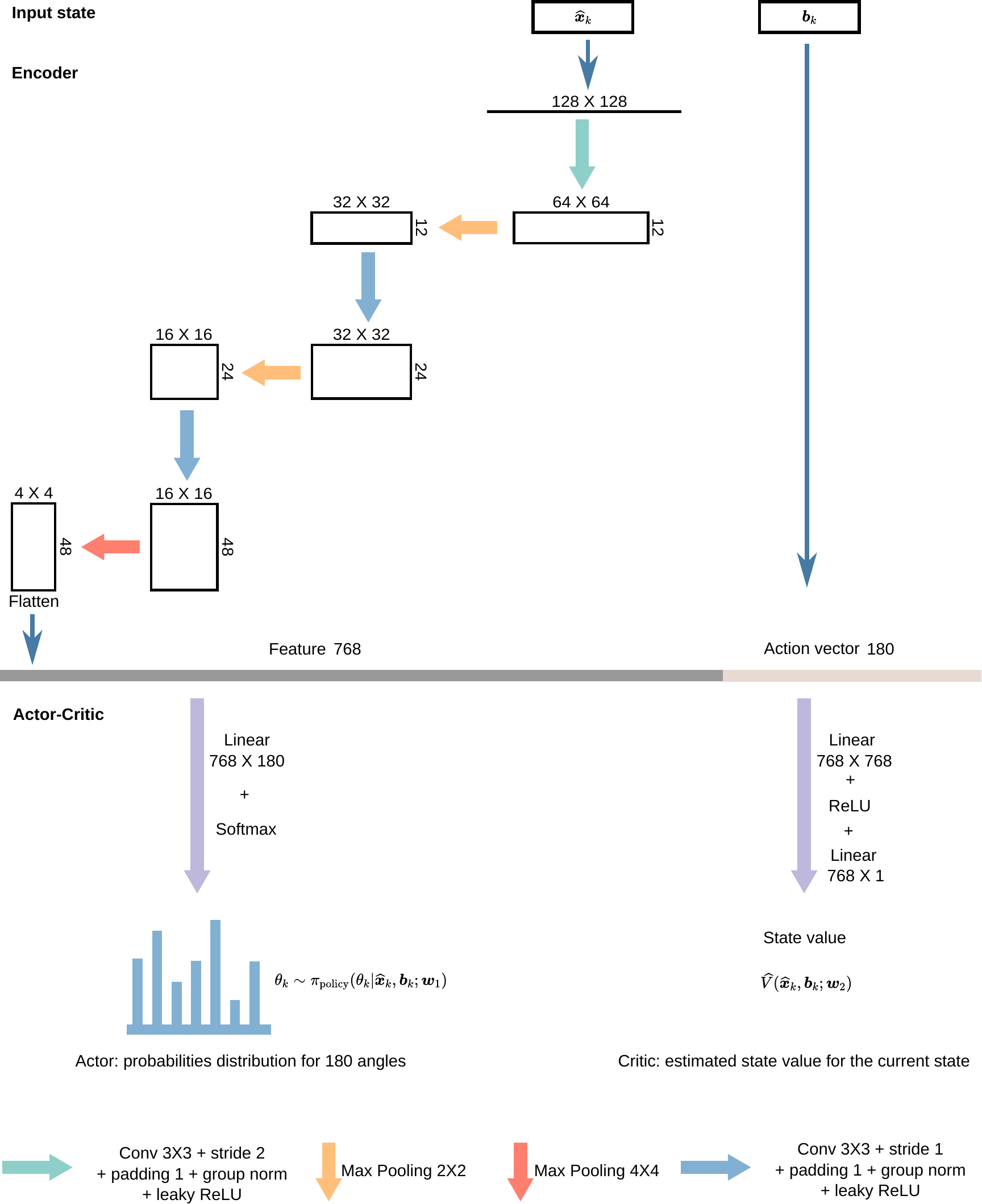}
  \caption{The combined network architecture first consists of an image encoder branch that processes the current reconstruction. Then, the code is concatenated with the previous action vector and fed into a network parameterizing the policy (Actor, left bottom) and a network estimating the state value (Critic, right bottom).}
  \label{fig:Network}
\end{figure}

The proposed method requires the agent to extract relevant features from high-dimensional images to increase learning efficiency, which is accomplished using a deep neural encoder network. The architecture of the encoder network and the Actor-Critic network is shown in Figure (\ref{fig:Network}), with the input image being of dimension $128 \times 128$. The neural network's connection weights represent the policy parameters $\boldsymbol{w}_{1}$ and the state-value function parameters $\boldsymbol{w}_{2}$.

The proposed model adopts a shared encoder network between the actor and critic networks. This shared encoder network comprises three convolutional neural networks (CNNs) each with padding and group normalization, followed by a leaky Rectified Linear Unit (ReLU) activation and a max pooling operation for down-sampling. The shared encoder network consists of a total of 13,320 parameters. Furthermore, the following actor and critic networks are separate and have 170,820 and 900,601 parameters, respectively.

Figure (\ref{fig:Network}) outlines the process by which the network operates in the context of the Actor-Critic method. The encoder network takes the reconstruction $\widehat{\boldsymbol{x}}_{k}$ as input and produces a feature vector in the bottleneck layer, which is flattened into a 1D vector and concatenated with the 1D action vector $\boldsymbol{b}_{k}$. The resulting information is then fed into the following actor and critic networks.

The actor network uses a Soft-max policy to map the information to a probability distribution over all possible angle candidates in the action space, while the critic network estimates the state-value function $\widehat{V}(\widehat{\boldsymbol{x}}_{k}, \boldsymbol{b}_{k};\boldsymbol{w}_{2})$. Based on the probability distribution generated by the actor network, the agent selects the next angle $\theta_{k}$ and subsequently collects measurements to obtain a new reconstruction $\widehat{\boldsymbol{x}}_{k+1}$. The action vector is updated as $\boldsymbol{b}_{k+1}$ accordingly.

To compute the policy gradient and update the parameters in the value function using TD error in Algorithm (\ref{alg:alg1}), the new reconstruction $\widehat{\boldsymbol{x}}_{k+1}$ and the new action vector $\boldsymbol{b}_{k+1}$ are fed into the network again. This is done to calculate a new state-value function $\widehat{V}(\widehat{\boldsymbol{x}}_{k+1}, \boldsymbol{b}_{k+1};\boldsymbol{w}_{2})$. Once an angle is selected, both the policy parameters $\boldsymbol{w}_{1}$ and the value function parameters $\boldsymbol{w}_{2}$ are updated once. 

\section{Numerical Experiments}

We examine in some numerical experiments whether the learned policies are really able to sequentially adapt the scan angles to the object (a-posteriori adaptation). For this, we use various simple numerical phantoms—where informative angles are well-established—as well as more realistic phantoms. Throughout our experiments, we focus on parallel-beam geometry and 2D tomography using synthetic data. The code and synthetic data are available on Github \footnote{\url{https://github.com/tianyuan1wang/SeqAngleRL}}.

\subsection{Datasets}
\begin{figure}[t]
  \centering
  \includegraphics[width=\linewidth]{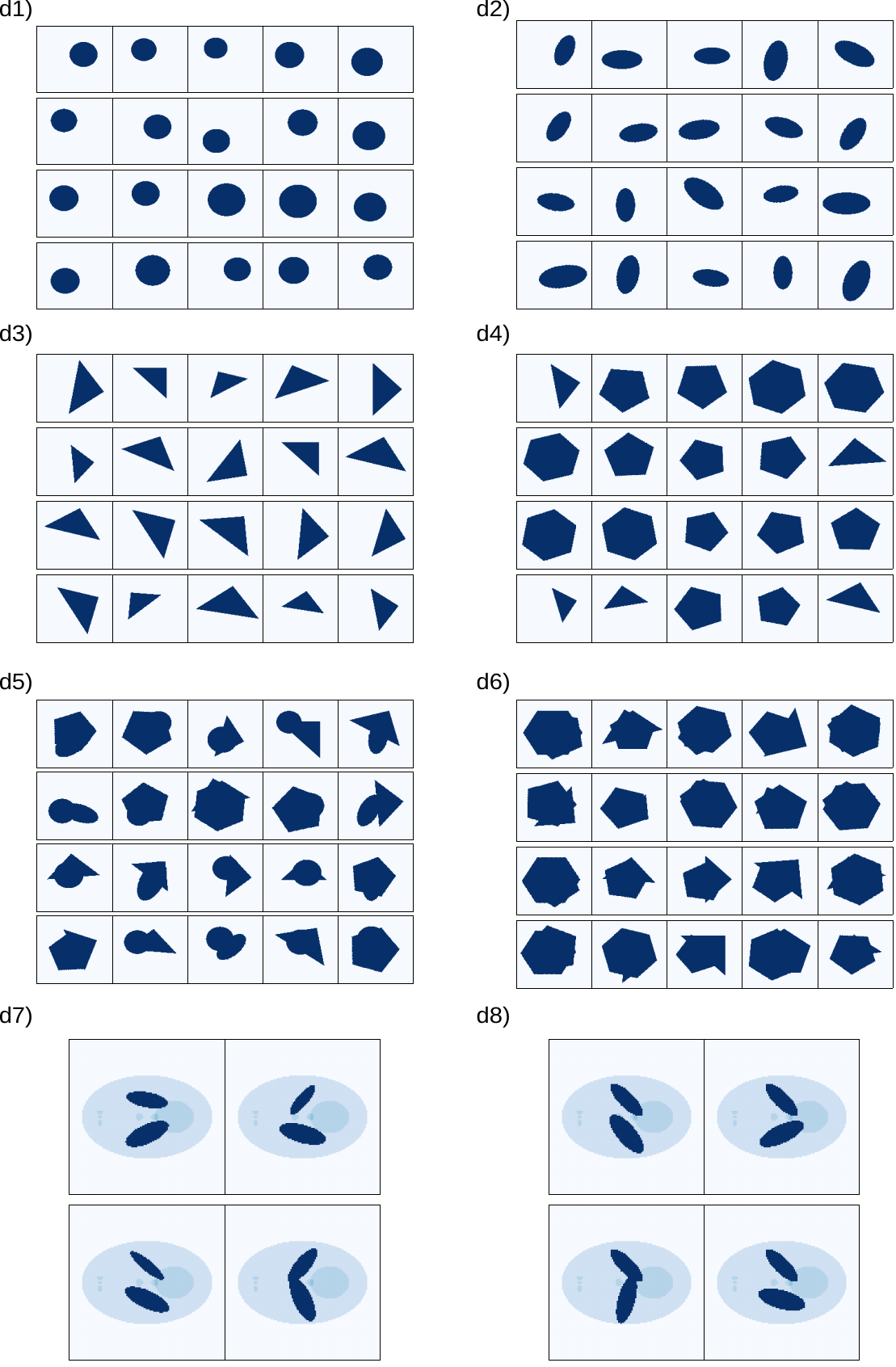}
  \caption{Datasets utilized in subsequent experiments: Each dataset comprises samples that exhibit a variety of rotations and scales.}
  \label{fig:Dataset}
\end{figure}

In our numerical experiments, we consider several shapes depicted in Figure (\ref{fig:Dataset}). All phantoms have a size of $128 \times 128$. The phantoms in the datasets from d1) to d5) are binary images. To facilitate algorithm validation with more sophisticated scenarios, datasets d7) and d8) feature advanced phantoms composed of multiple materials. To assess the adaptability of the agent to dynamic environments, each dataset from d1) to d8) includes phantoms with different rotations, causing a shift in their informative angles. These rotations are represented by 36 equally spaced angles ranging from $0^{\circ}$ to $179^{\circ}$. Additionally, the phantoms in each dataset from d1) to d8) exhibit various scaling and shifts (parameters are shown in Appendix \ref{appendix:A}). Nonetheless, these modifications do not alter the informative angles, thereby preserving the consistency of informative angles across the scaled and shifted phantoms. By including these scaling and shifts, we aim to ensure the agent's ability to recognize the same object despite its size and location variations. Using these datasets allows us to transparently validate the algorithm.

\textbf{d1) Circles}: The first dataset consists of circles with varying locations and radii. Due to its uniform curvature, a circle does not have a relatively higher concentration of informative angles. To obtain an accurate reconstruction, angles must be equidistantly distributed. 

\textbf{d2) Ellipses}: Unlike circles, ellipses have a major axis and a minor axis. The major axis serves as a preferential direction, making angles around it more informative, as shown in references \cite{batenburg2013dynamic} and \cite{venere2000genetic}. 

\textbf{d3) Triangles}: Triangles, characterized by one angle of $90^{\circ}$ and two angles of $45^{\circ}$, possess three preferential directions, causing the informative angles to be tangential to their edges \cite{yang2023edge}.

\textbf{d4) Mixed phantoms}: This dataset consists of a mixture of phantoms, including triangles from d3), regular pentagons, and regular hexagons, each of which has its own preferential directions.

\textbf{d5) and d6) Combined phantoms}: The dataset d5) comprises samples created by combining elements from two distinct phantom types found in categories d1), d2), and d4). In contrast, dataset d6) consists of two distinct samples from dataset d4), which does not include round shapes. These two datasets integrate features from a variety of shapes. Through this approach, the dataset transforms simple phantoms into more complex entities, facilitating a broader exploration of shape and structure.

\textbf{d7) and d8) Complex phantoms}: The datasets under consideration introduce variations to the classic Shepp-Logan phantoms. Specifically, we have redesigned the Shepp-Logan phantoms by excluding the outer ellipse and modifying the two inner ellipses, ranging from the lowest to the highest intensity values. In contrast to previous datasets, the current ones integrate both static and dynamic elements. In the d7) setup, two high-intensity ellipses experience changes in both scale and rotation across 36 uniformly distributed angles from $0^{\circ}$ to $179^{\circ}$. Following such modifications, these redefined ellipses markedly impact the reconstruction error, consequently influencing the adaptive capabilities of the trained policy. On the other hand, in the d8) setup, only the lower of these ellipses exhibits dynamic characteristics.

\subsection{Implementation}
For all of our experiments, the sequential experimental process for each dataset in Figure (\ref{fig:Dataset}) follows Algorithm (\ref{alg:alg1}).  To generate the measurement data, we utilize the ASTRA Toolbox \cite{van2015astra, van2016fast}, considering a projection size of $1.5 \times 128$. This specific size choice ensures that each angle covers all pixels. The reconstruction is performed using the SIRT algorithm with box constraints [0,1] for 150 iterations.

The encoder and Actor-Critic neural network architectures are illustrated in Figure (\ref{fig:Network}). During training, we set the discount factor $\gamma$ to 0.99 and assign weights of 1.0 and 0.5 to the actor loss: $-\log \pi_{\text{policy}}(\theta_{k}|\widehat{\boldsymbol{x}}_{k}, \boldsymbol{b}_{k};\boldsymbol{w}_{1}) \delta_{k}$ and critic loss: $(\delta_{k})^{2}$, respectively. To encourage exploration during training, we incorporate an entropy loss: $\sum\limits_{\theta}\pi_{\text{policy}}(\theta_{k}|\widehat{\boldsymbol{x}}_{k}, \boldsymbol{b}_{k};\boldsymbol{w}_{1}) \log \pi_{\text{policy}}(\theta_{k}|\widehat{\boldsymbol{x}}_{k}, \boldsymbol{b}_{k};\boldsymbol{w}_{1})$, with a weight of 0.01. The aforementioned parameter settings have been empirically determined to achieve an optimal balance between policy optimization, accurate value estimation, and robust exploration throughout the training regimen. For optimization of the parameters, we employ the Adam optimizer \cite{kingma2014adam} with a learning rate of $10^{-4}$ and weight decay of $10^{-5}$.

In addition, we consider two reward functions: incremental and end-to-end settings. An equidistant policy is introduced as a benchmark to compare the performance of the Actor-Critic policy with un-informed and non-adaptive angle selection method.  Significantly, Experiments 1, 2, 3, 4, and 6 employ distinct policies, each specifically tailored to their respective training datasets. Conversely, Experiment 5 adopts either the policy developed in Experiment 4 or a newly trained policy encompassing multiple datasets, to thoroughly evaluate and validate its ability to generalize.

\subsection{Experiment 1 - Uniform informative angles}
In the first experiment, we aim to train an agent and evaluate its performance on the circles (dataset d1), which have a uniform distribution of informative angles. It is known that the equidistant benchmark is the optimal policy for this dataset. Our objective is to investigate whether the Actor-Critic policy approaches the equidistant policy in performance. To facilitate this investigation, the experiment is conducted with 3,000 training phantoms, and it requires an extensive training duration of 100,000 episodes.

\begin{figure}[t]
  \centering
  \includegraphics[width=\linewidth]{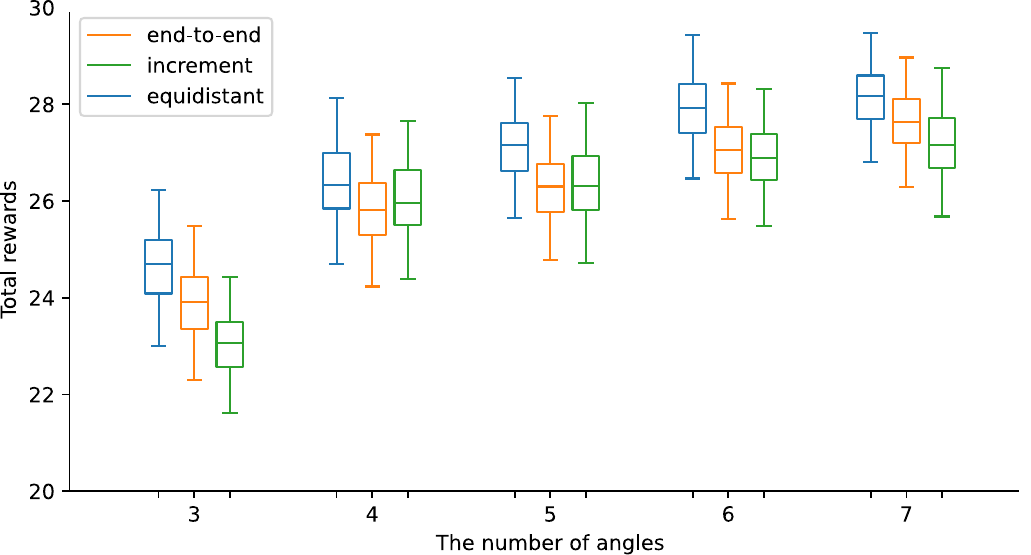}
  \caption{Comparison of policies considering different numbers of angles for the circles dataset: the results demonstrate the training outcomes over the last 2,000 episodes. The box represents the interquartile range in these plots, spanning from the first to the third quartile of the data distribution. The median value is displayed as a line within the box. The whiskers extend from the box to illustrate the range of the data distribution beyond the interquartile range.}
  \label{fig:CirleTraining}
\end{figure}

As depicted in Figure (\ref{fig:CirleTraining}), the equidistant policy exhibits enhanced performance for the circular phantoms compared to the Actor-Critic policies with diverse reward configurations. Furthermore, we observe that the performance of the Actor-Critic policy with end-to-end reward surpasses that of the policy with incremental reward as the number of angles increases.

\begin{figure}[t]
  \centering
  \includegraphics[width=\linewidth]{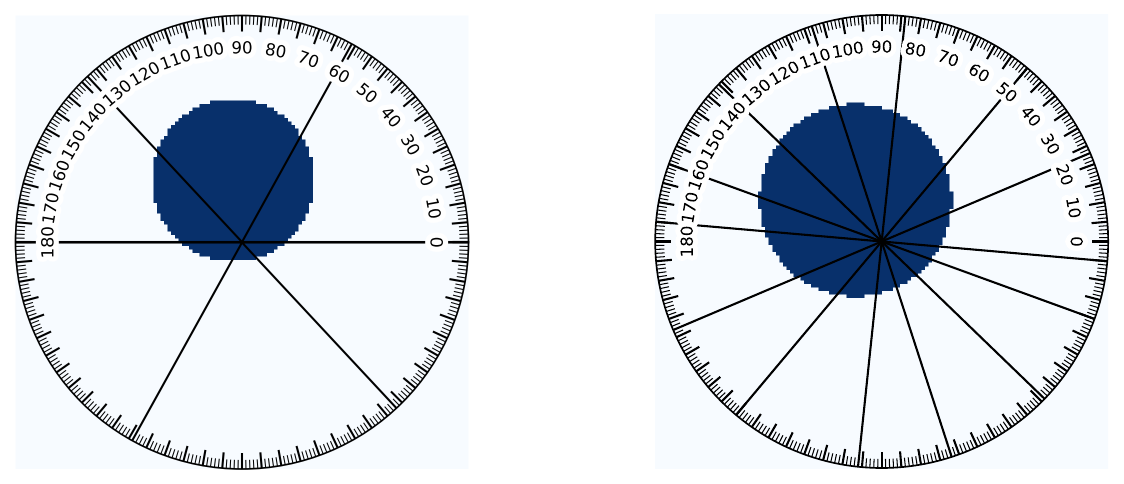}
  \caption{Results of the end-to-end reward setting for two circle phantoms considering three and seven angles.}
  \label{fig:CircleSamples}
\end{figure}

Figure (\ref{fig:CircleSamples}) presents two samples considering three and seven angles obtained from the Actor-Critic policy with the end-to-end reward setting. This result demonstrates that the Actor-Critic agent tends to distribute the selected angles evenly, although the number of angles is different. 

\subsection{Experiment 2 - Non-uniform informative angles}
In contrast to circles, informative angles in ellipses (dataset d2) are found to be concentrated around its major axis. We train an agent on a dataset comprising 3,000 ellipse phantoms. The model achieves convergence after undergoing 150,000 training episodes.

\begin{figure}[t]
  \centering
  \includegraphics[width=\linewidth]{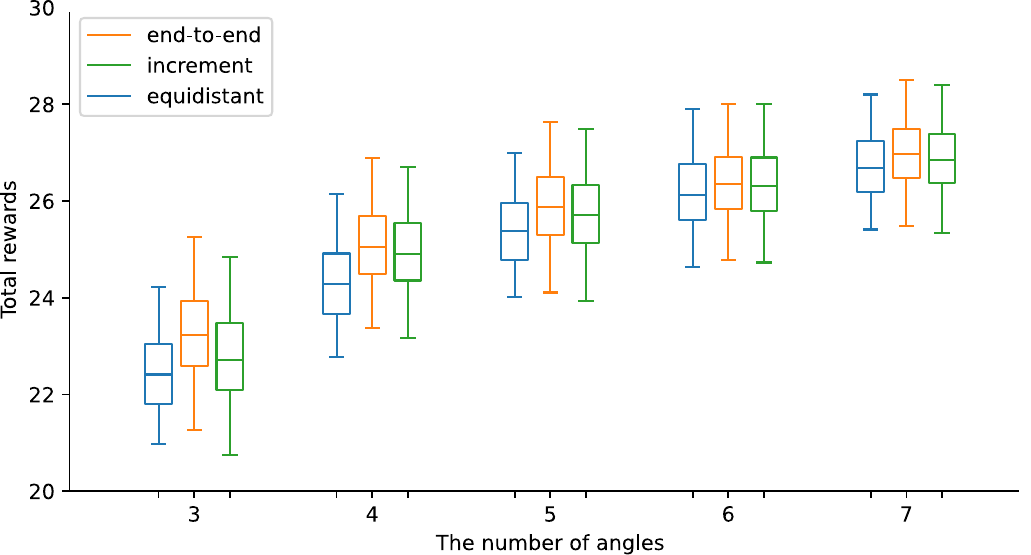}
  \caption{Comparison of policies considering different numbers of angles for the ellipses dataset: the results demonstrate the training outcomes over the last 2,000 episodes. The box represents the interquartile range in these plots, spanning from the first to the third quartile of the data distribution. The median value is displayed as a line within the box. The whiskers extend from the box to illustrate the range of the data distribution beyond the interquartile range.}
  \label{fig:EllipseTraining}
\end{figure}

\begin{figure}[t]
  \centering
  \includegraphics[width=\linewidth]{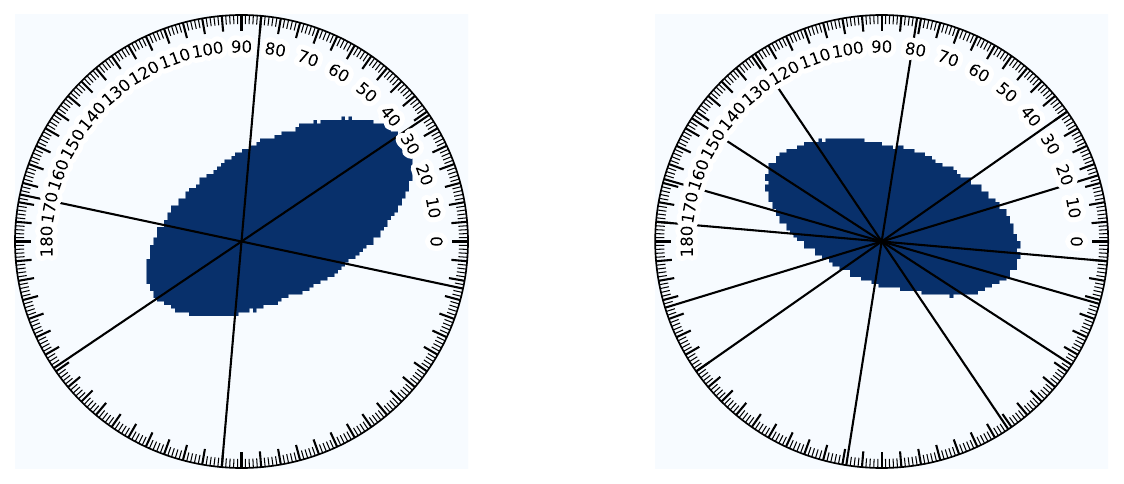}
  \caption{Results of the end-to-end reward setting for two ellipse phantoms considering three and seven angles.}
  \label{fig:EllipseSamples}
\end{figure}

The training outcomes of the ellipse phantoms over the final 2,000 episodes are shown in Figure (\ref{fig:EllipseTraining}), which indicates that the Actor-Critic policies exhibit superior performance. As the number of angles increases, the results for the three policies get closer. This is because a sufficient number of angles around the major axis have already been obtained, even for the equidistant policy, to achieve a high-quality reconstruction. Notably, the Actor-Critic policy with the end-to-end reward setting achieves the best performance. Figure (\ref{fig:EllipseSamples}) presents the results for two ellipse phantoms, demonstrating that the agent can discern the rotation of the ellipse and concentrate the distribution of the angles around the informative area. As the number of angles increases, the agent increases the number of angles around the major axis.

The unseen test set comprises 300 phantoms. Table (\ref{tb:EllipseTest}) reports the outcomes regarding the unseen rotations for the ellipses dataset with three to seven angles. Consistent with the training results, the Actor-Critic policies demonstrate superior performance compared to the benchmark, with the policies becoming progressively closer as the number of angles increases. The end-to-end reward setting still shows the best average performance, though it has a higher variance.

\begin{table*}[!t]
\caption{Performance comparison of policies on unseen rotations test for ELLIPSES regarding the PSNR values}
\centering
\begin{tabular}{c|c|c|c|c|c}

\textbf{Policies} & \textbf{3} & \textbf{4} & \textbf{5} & \textbf{6} &  \textbf{7}\\ 
\hline
Learned adaptive policy (end-to-end) & \bfseries 23.16 $\pm$ 1.02 & \bfseries 25.10 $\pm$ 0.72 & \bfseries 25.73 $\pm$ 1.20 & 26.31 $\pm$ 0.82 & \bfseries 26.87 $\pm$ 1.06 \\

Learned adaptive policy (increment) & 22.78 $\pm$ 0.92 & 24.90 $\pm$ 0.73 & 25.67 $\pm$ 0.77 & \bfseries 26.33 $\pm$ 0.79 & 26.86 $\pm$ 0.73 \\

Equidistant policy & 22.40 $\pm$ 0.74 & 24.27 $\pm$ 0.73 & 25.35 $\pm$ 0.69 & 26.16 $\pm$ 0.64 & 26.73 $\pm$ 0.62 \\

\label{tb:EllipseTest}
\end{tabular}

\end{table*}

\subsection{Experiment 3 - Explicit informative angles}
The third experiment focuses on evaluating the ability of the Actor-Critic agent to identify explicit informative angles for phantoms with sharp edges, namely triangles (dataset d3). These phantoms have well-defined informative angles that are tangential to their edges, and thus, it is of interest to investigate if the agent can successfully locate these angles. The results of this experiment will provide insight into the performance of the Actor-Critic agent in detecting preferential directions for phantoms with sharp edges. 

\begin{figure}[t]
  \centering
  \includegraphics[width=\linewidth]{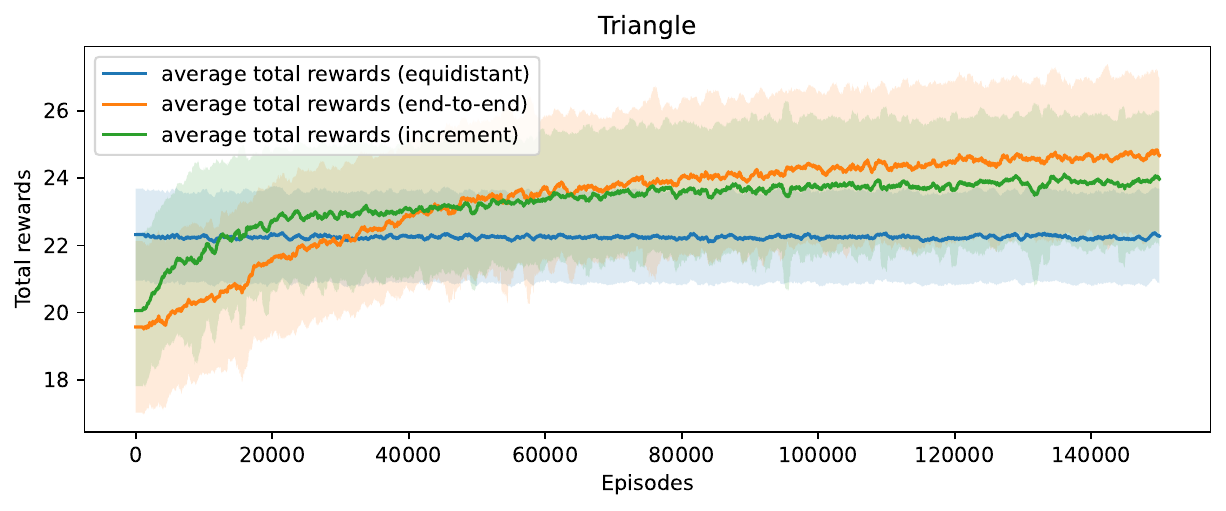}
  \caption{This figure compares the performance of Actor-Critic policies, trained on triangles phantoms, with that of an equidistant policy. It displays the training outcomes, with curves showing the mean values and shaded color bands representing the variances.}
  \label{fig:TriangleTraining}
\end{figure}

This study employs a dataset comprising 3,000 triangle phantoms for training the agent. A total of 150,000 episodes are required to reach convergence in the training process. A fixed number of five angles is employed. As shown in Figure (\ref{fig:TriangleTraining}), both reward settings for the Actor-Critic agent outperform the equidistant policy. Specifically, training using the incremental reward setting demonstrates faster convergence, whereas the end-to-end reward setting yields the best performance.

The training results demonstrate that the Actor-Critic agent tends to select the first two angles as fixed angles, with particular emphasis on the first angle, while the second angle exhibits some uncertainty. Subsequently, the agent would select three informative angles to optimize the reconstruction process. This behavior is consistent with the fact that the initial state is set as a zero matrix and a zero vector with no prior information, and the agent, therefore, prioritizes gathering information by fixing the first angle or first two angles before personalizing the strategies based on the different phantoms encountered.

\begin{figure}[t]
  \centering
  \includegraphics[width=\linewidth]{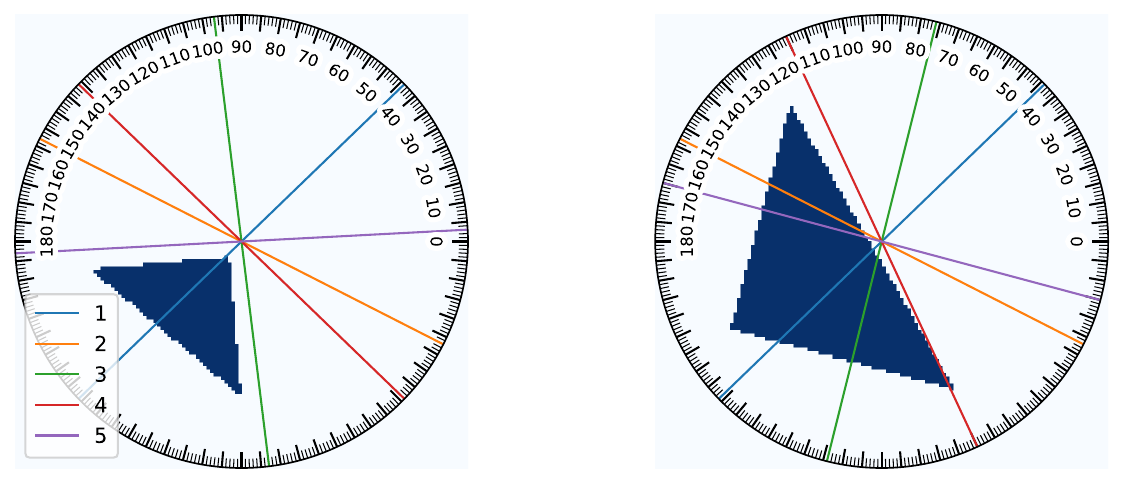}
  \caption{The personalized strategies for triangle phantoms achieved by the Actor-Critic policy are demonstrated in these sample results obtained under the end-to-end reward setting. }
  \label{fig:TriangleSamples}
\end{figure}

Figure (\ref{fig:TriangleSamples}) presents two samples of the agent's performance in an end-to-end reward setting, in which the agent selects the initial two angles of $44^{\circ}$ and $153^{\circ}$. Subsequently, for the right phantom, the agent selects $76^{\circ}$, $115^{\circ}$, and $165^{\circ}$ as the following three angles, while for the left phantom, the agent chooses $97^{\circ}$, $136^{\circ}$, and $3^{\circ}$. Notably, these angles are almost tangential to the edges of the triangle phantoms. 

We observe that sometimes, the agent selects more angles around the informative angles or repeats its selection when the first two angles are close to the informative angles. This sub-optimal behavior is counter-intuitive and might indicate that the policy network's capacity is too limited to avoid this repetition. 

\begin{table}[!t]
\caption{Performance comparison of policies on unseen rotations test for TRIANGLES regarding the PSNR values}
\centering
\begin{tabular}{c|c}

\textbf{Policies} & \textbf{Triangles}\\ 
\hline
Learned adaptive policy (end-to-end) & \bfseries 24.07 $\pm$ 2.07  \\

Learned adaptive policy (increment) & 23.78 $\pm$ 1.80 \\

Equidistant policy & 20.64 $\pm$ 1.05  \\

\label{tb:TriangleTest}
\end{tabular}
\end{table}

In the context of evaluating unseen rotations across 300 test phantoms, Table (\ref{tb:TriangleTest}) demonstrates that the Actor-Critic policies outperform the equidistant benchmark. Furthermore, it is observed that the end-to-end reward setting achieves the highest quality in terms of reconstruction.

\subsection{Experiment 4 - Mixed phantoms with explicit informative angles}
In this study, we aim to investigate the capacity of an Actor-Critic agent to recognize and distinguish between different phantoms with varying informative angles and their rotation. Our research methodology trains the agent using a comprehensive dataset (dataset d4) comprising 9,000 phantoms over 300,000 training episodes. In this study, a fixed number of seven angles is employed. 

\begin{figure}[t]
  \centering
  \includegraphics[width=\linewidth]{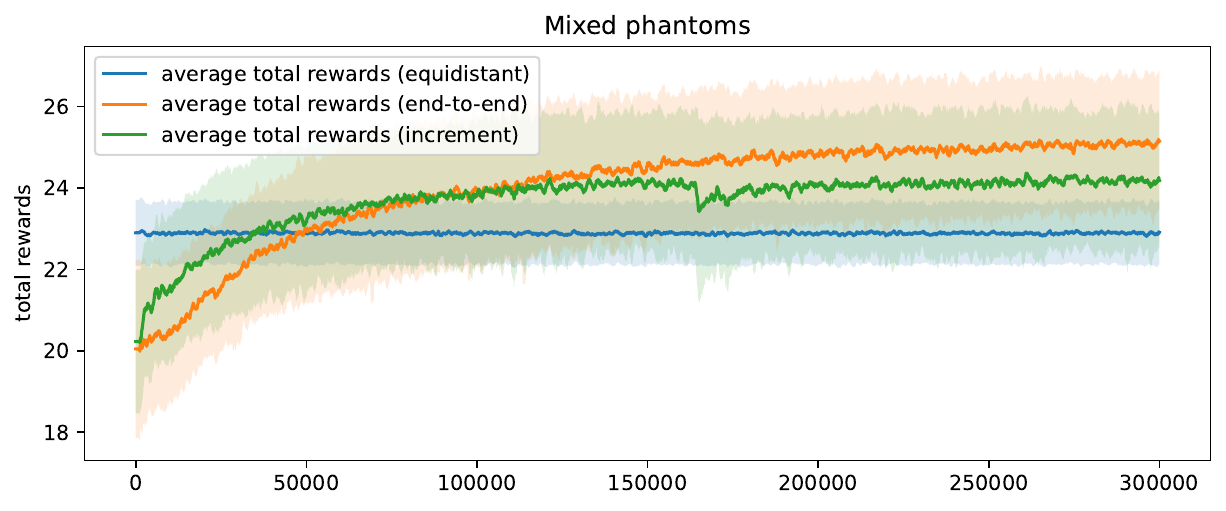}
  \caption{This figure compares the performance of Actor-Critic policies, trained on mixed phantoms dataset, with that of an equidistant policy. It displays the training outcomes, with curves showing the mean values and shaded color bands representing the variances.}
  \label{fig:MixedTraining}
\end{figure}

\begin{figure*}[t]
  \centering
  \includegraphics[width=\linewidth]{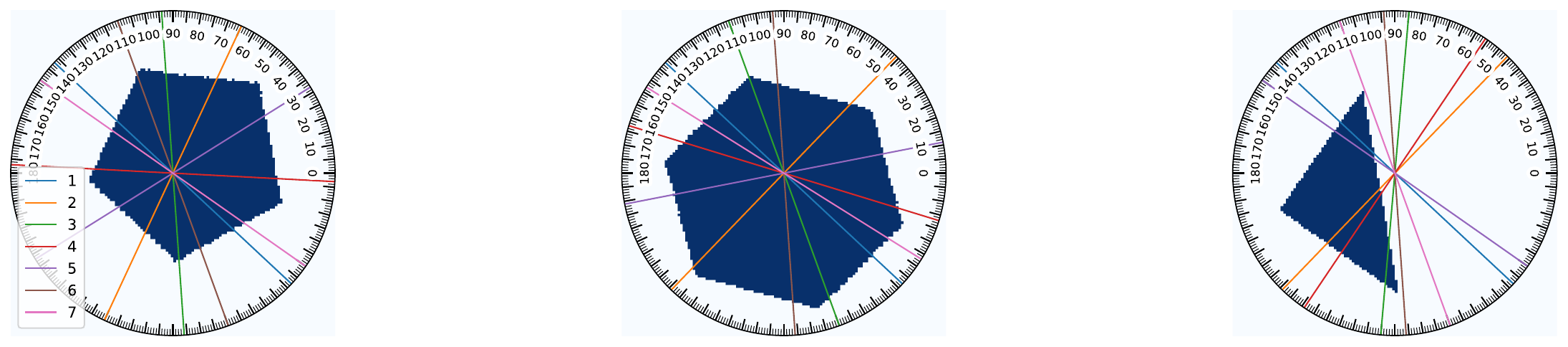}
  \caption{The personalized strategies for mixed phantoms achieved by the Actor-Critic policy are demonstrated in these sample results obtained under the end-to-end reward setting. }
  \label{fig:MixedSamples}
\end{figure*}

Similar to Experiment 3, shown in Figure (\ref{fig:MixedTraining}), the training for the mixed phantoms reveals that the incremental reward setting facilitates faster convergence, while end-to-end reward setting results in better performance. Figure (\ref{fig:MixedSamples}) illustrates the performance of the end-to-end reward setting. It fixes the first two angles to $137^{\circ}$ and $46^{\circ}$ for the hexagon and triangle, respectively, while it selects $137^{\circ}$ and $65^{\circ}$ for the pentagon as the first two angles because of some uncertainty for the second angle as mentioned in Experiment 3. The agent then selects the subsequent informative angles based on this prior information by the fixed angles. The reconstruction results for these samples are illustrated in Appendix \ref{appendix:B}.

\begin{table}[!t]
\caption{Performance comparison of policies on unseen rotations test for MIXED phantoms regarding the PSNR values}
\centering
\begin{tabular}{c|c}

\textbf{Policies} & \textbf{Mixed phantoms}\\ 
\hline
Learned adaptive policy (end-to-end) & \bfseries 24.85 $\pm$ 1.64  \\

Learned adaptive policy (increment) & 24.15 $\pm$ 1.55 \\

Equidistant policy & 22.94 $\pm$ 0.76  \\

\label{tb:MixedTest}
\end{tabular}
\end{table}

Drawing from the unseen rotations test conducted on 900 phantoms, as detailed in Table (\ref{tb:MixedTest}), it can be observed that the Actor-Critic policies, with both end-to-end and incremental rewards, outperform the equidistant policy. 

\subsection{Experiment 5 - Generalizability Assessment}
To evaluate the generalizability of the proposed method, we apply the trained Actor-Critic policy, initially tailored for mixed phantoms, to various noise levels, phantoms possessing distinct shapes and intensity values. Following this, we train the Actor-Critic policy using simple datasets. Next, to assess its effectiveness, we evaluated how well it could generalize from a simple training dataset to a similar yet more complex dataset.

\begin{figure}[t]
  \centering
  \includegraphics[width=\linewidth]{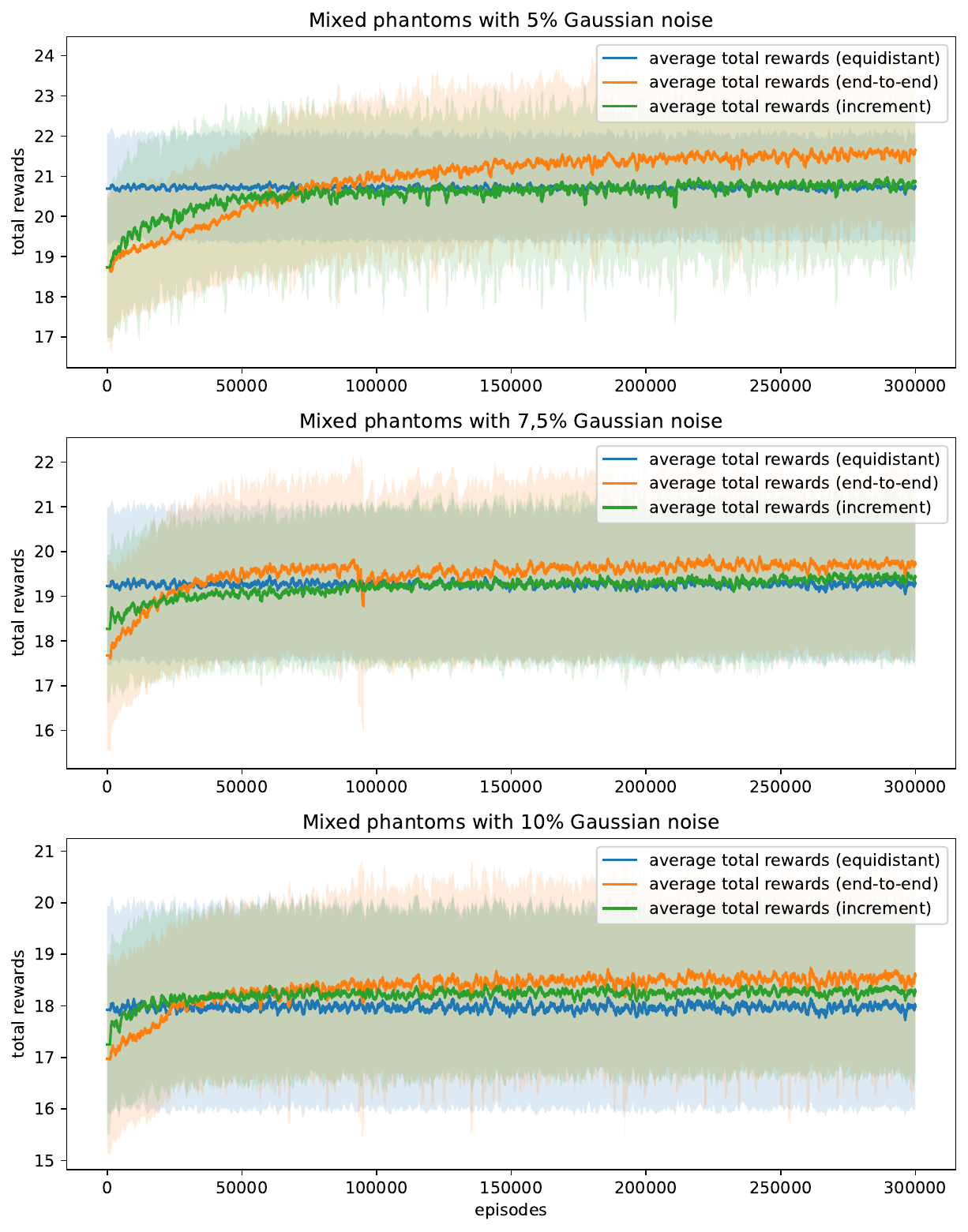}
  \caption{This figure compares the performance of Actor-Critic policies, trained on a mixed phantoms dataset, with that of an equidistant policy. It displays the training outcomes while considering the impact of Gaussian noise (5\%, 7.5\%, and 10\%), with curves indicating the mean values and shaded color bands representing the variances.}
  \label{fig:MixedTrainingN}
\end{figure}

\begin{figure*}[t]
  \centering
  \includegraphics[width=\linewidth]{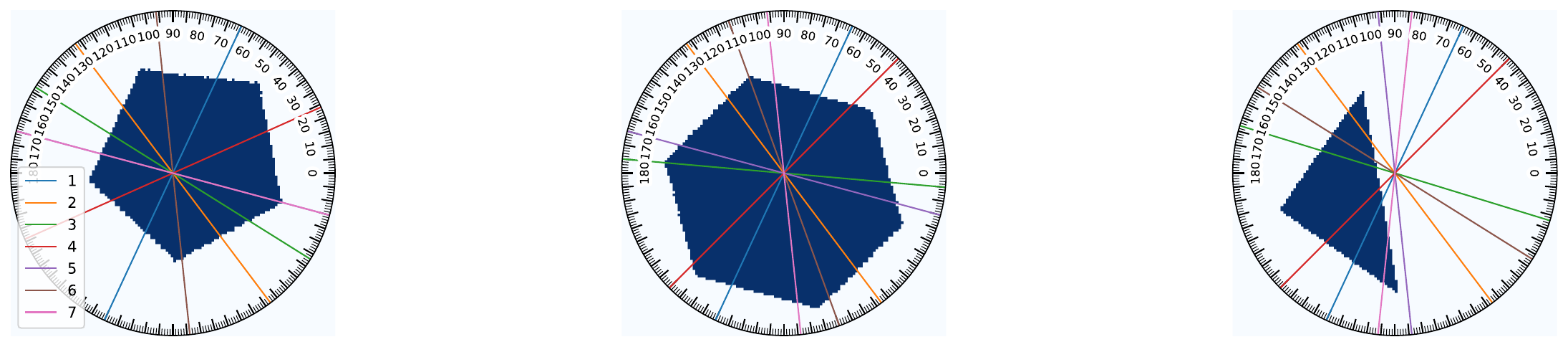}
  \caption{The personalized strategies for mixed phantoms with $5\%$ Gaussian noise on measurements achieved by the Actor-Critic policy are demonstrated in these sample results, obtained under the end-to-end reward setting.}
  \label{fig:MixedNSamples}
\end{figure*}

\begin{figure*}[t]
  \centering
  \includegraphics[width=\linewidth]{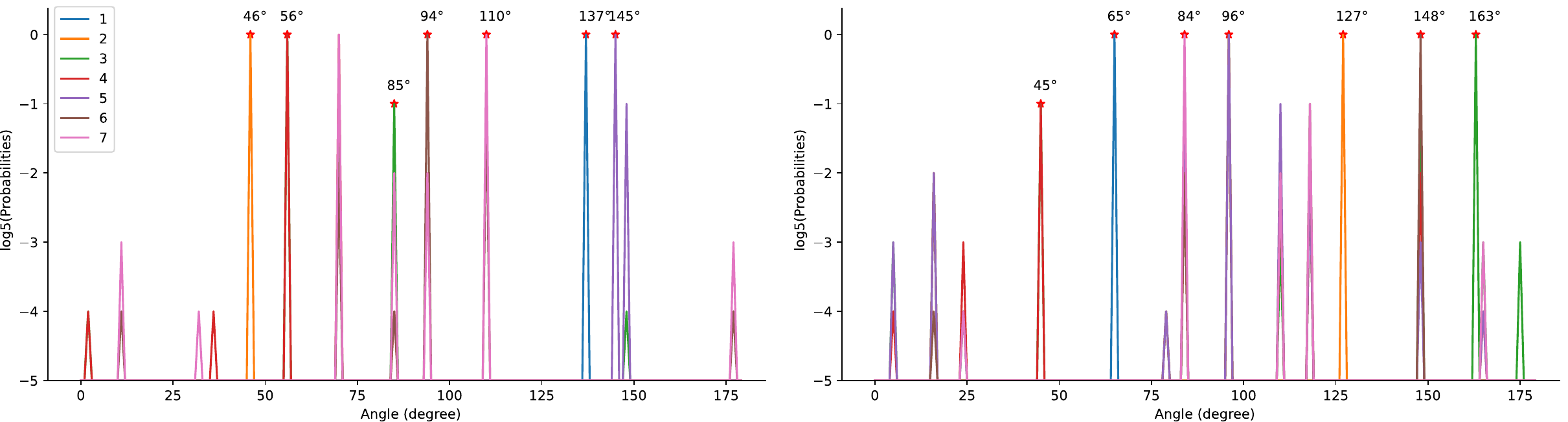}
  \caption{Comparison of policies for triangles. The policy on the left is trained on clean data, while the policy on the right is trained on noisy data. The probabilities are scaled by $\log5$ to magnify the small probabilities. The red star on each colored line represents the selected angle based on the corresponding probability distribution. Once an angle is selected, its probability becomes zero or is decreased for subsequent selections.}
   \label{fig:ComparisonPolicy}
\end{figure*}

We investigate the impact of Gaussian noise ($5\%$, $7.5\%$, and $10\%$) on the Actor-Critic agent's ability to select informative angles for phantoms. As illustrated in Figure (\ref{fig:MixedTrainingN}), the incorporation of Gaussian noise into measurements results in a decrease in performance for both the equidistant and Actor-Critic policies, particularly when compared to outcomes from noise-free measurements. Nevertheless, Actor-Critic policies consistently outperform the equidistant approach, particularly in an end-to-end framework. This indicates that the predominant factors contributing to performance disparities are the difficulties inherent in reconstructing data from noise-contaminated inputs. Comparing the result samples with 5$\%$ Gaussian noise from end-to-end rewards in Figure (\ref{fig:MixedNSamples}) to those in Figure (\ref{fig:MixedSamples}) from Experiment 4, we find that the noise in measurements has a substantial influence on the angle selection strategy, including the fixed angles and the informative angles selection orders afterward. To better understand the differences between the two policies in Figures (\ref{fig:MixedSamples}) and (\ref{fig:MixedNSamples}), we show the policy results for triangles in Figure (\ref{fig:ComparisonPolicy}). Our analysis reveals that the policy for clean data is tightly clustered around informative angles, whereas the policy for noisy data is more broadly distributed. We also observe that the policy realizes adaptive angle selection, where the probability of a chosen angle decreases significantly, followed by an increase in the probability of some angles with a small probability. 

\begin{figure}[t]
  \centering
  \includegraphics[width=\linewidth]{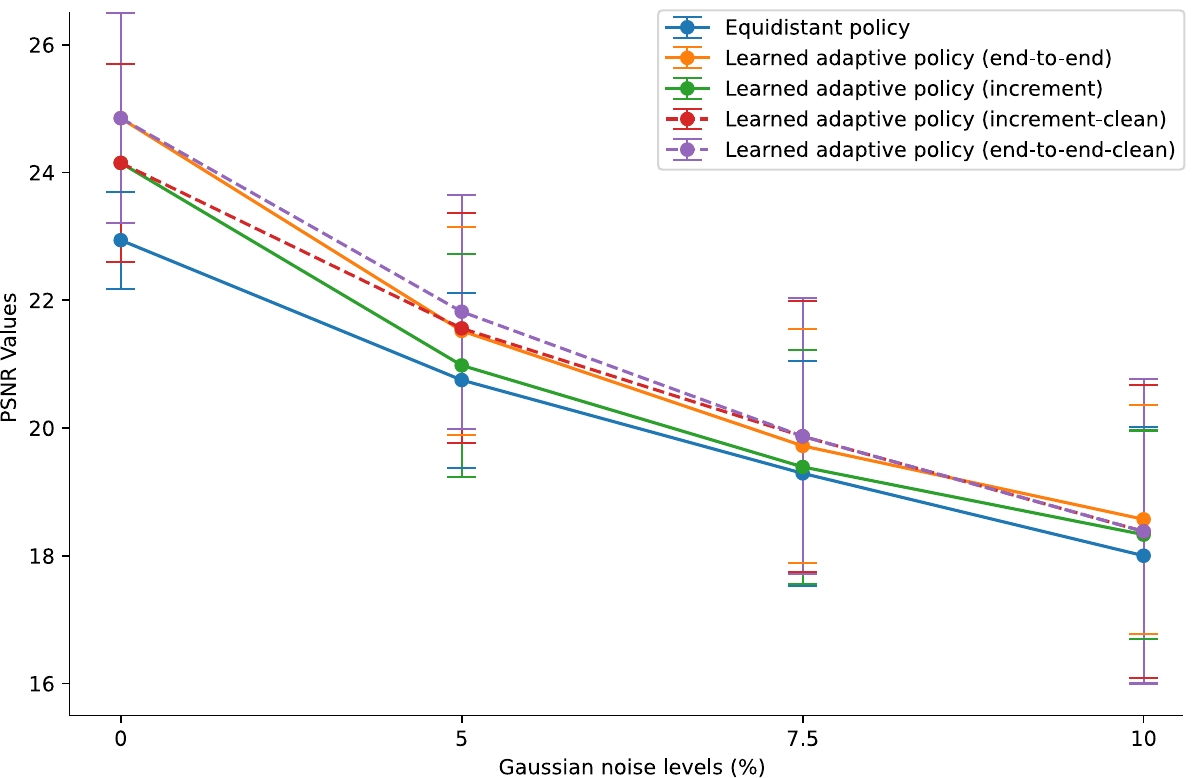}
  \caption{Comparison of policies considering different Gaussian noise levels for the mixed dataset d4): the results demonstrate the unseen rotations outcomes over. The graph depicts the efficacy of learned adaptive policies, both in incremental and end-to-end settings, against the equidistant policy. Notably, both clean and noisy training conditions are considered for the adaptive policies. The x-axis represents the Gaussian noise percentage, while the y-axis captures the performance metric. Error bars indicate the variability in performance at each noise level.}
  \label{fig:CompareNoise}
\end{figure}

In the test involving unseen rotations, conducted on 900 phantoms with variable noise levels as depicted in Figure (\ref{fig:CompareNoise}), the Actor-Critic policies consistently surpass the equidistant policy, with the end-to-end reward setting maintaining the most superior performance across the three distinct noise levels. This suggests again that the observed performance variation predominantly stems from the reconstruction challenges posed by noisy measurements rather than the efficiency of the trained policy itself. Additionally, models trained on clean measurements generally exhibit enhanced resilience to noisy measurements compared to those trained on noisy datasets, suggesting that training with noisy data might hinder the optimal performance of the Actor-Critic policies.

In terms of shape variations, we analyze phantoms with fewer edges, such as rectangles, and those with a greater number of edges, specifically heptagons, and octagons. For each shape category, we have studied a sample size of 900 phantoms. As deduced from Table (\ref{tb:ShapesTest}), while the Actor-Critic policy demonstrates a degree of generalizability for diverse phantom shapes, such as heptagon, its efficacy is somewhat restricted. The policy with an incremental setting demonstrates enhanced generalizability.

\begin{table*}[!t]
\caption{Performance comparison of policies on test for phantoms with different shapes regarding the PSNR values}
\centering
\begin{tabular}{c|c|c|c}
\textbf{Policies} & \textbf{Rectangular} &   \textbf{Heptagon}  &  \textbf{Octagon}\\ 
\hline
Learned adaptive policy (end-to-end) &  21.28 $\pm$ 2.66 &  23.08 $\pm$ 0.74 & 23.19 $\pm$ 1.24\\

Learned adaptive policy (increment) & 21.89 $\pm$ 2.52 &  \bfseries 23.32 $\pm$ 0.57  & 23.46 $\pm$ 0.90\\

Equidistant policy & \bfseries 23.66 $\pm$ 1.20 &  22.75 $\pm$ 2.16 &  \bfseries 24.24 $\pm$ 0.43 \\

\label{tb:ShapesTest}
\end{tabular}
\end{table*}

For variations in intensity values, we assess phantoms (dataset d4) with four distinct intensity ranges: [0.9,1), [0.7,1), [0.5,1), and [0.1,1). For each of these intensity categories, our study encompassed a sample size of 900 phantoms. Insights drawn from Table (\ref{tb:IntensityTest}) suggest that as the intensity value range expands, the generalizability of the Actor-Critic policy diminishes.  However, the policy employing an end-to-end setting showcases superior generalizability.

\begin{figure}[t]
  \centering
  \includegraphics[width=\linewidth]{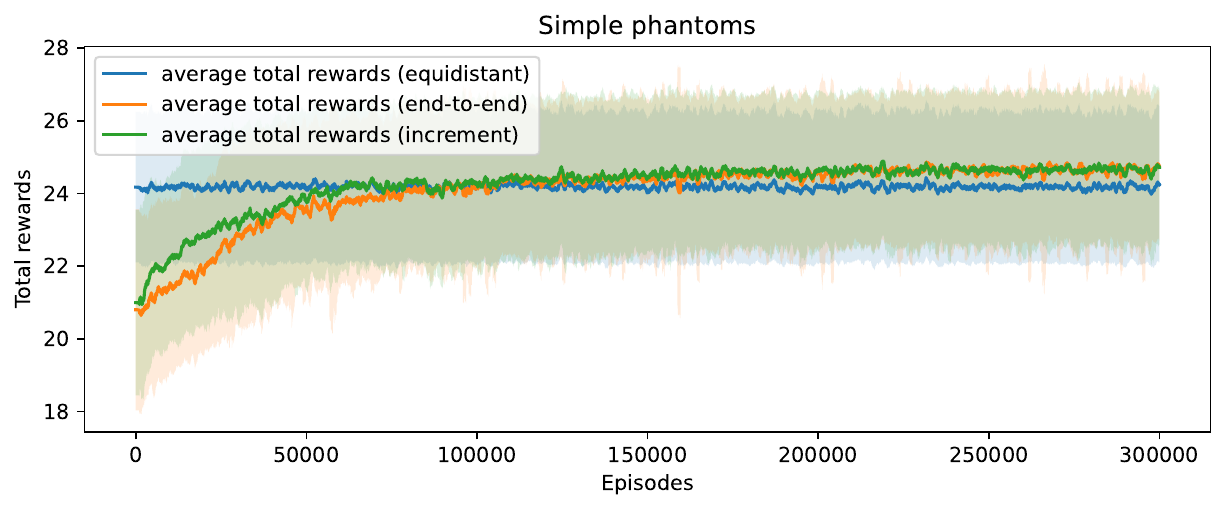}
  \caption{This figure compares the performance of Actor-Critic policies, trained on simple phantoms from datasets d1), d2), and d4), with that of an equidistant policy. It displays the training outcomes, with curves showing the mean values and shaded color bands representing the variances.}
  \label{fig:CombinedTraining}
\end{figure}

Finally, we initially trained the Actor-Critic policy on a comprehensive dataset that includes all phantoms from categories d1), d2), and d4). This approach aimed to develop a versatile and adaptable policy capable of handling a variety of shapes and complexities. With a set configuration of seven angles across 300,000 training episodes, Figure \ref{fig:CombinedTraining} showcases the training results for the Actor-Critic policies and uses the equidistant policy as a comparative baseline. The Actor-Critic policies surpass the equidistant policy in performance, though the margin is modest. This narrow performance gap can be attributed to the specific nature of the training datasets, particularly the circle and ellipse phantoms. Circles possess uniformly informative angles, while ellipses offer nearly uniform informative angles, as shown in Experiment 2, considering the seven angles. Interestingly, the performances of Actor-Critic policies, whether configured for end-to-end or incremental learning, show a remarkable similarity.

Subsequently, to assess the ability of the policy to generalize from a simpler training set to a more complex one, we evaluate the trained Actor-Critic policy on datasets d5) and d6) with a size of 900 phantoms, respectively. This category encompasses a mix of two distinct shape types found in the training dataset, thereby providing a challenging test for its adaptability. The results for the dataset d5), depicted in Table (\ref{tb:CombinedTest_d5}), indicate that the performance of these three policies is quite similar due to the inclusion of round shapes. In contrast, the results for dataset d6), shown in Table (\ref{tb:CombinedTest_d6}), demonstrate the superiority of both Actor-Critic policies. These results highlight the Actor-Critic’s proficiency in handling increased complexity. Additionally, the selection of angles for two specific samples, as detailed in Appendix \ref{appendix:E}, exemplifies the ability of the Actor-Critic policy to generalize from basic geometric shapes to more intricate, combined phantoms.

\begin{table}[!t]
\caption{Policy Performance on Complex Combined Phantoms: A comparison of policies trained on simple phantoms, evaluated on complex combined phantoms (d5) using PSNR values.}
\centering
\begin{tabular}{c|c}

\textbf{Policies} & \textbf{Combined phantoms d5)}\\ 
\hline
Learned adaptive policy (end-to-end) & 23.27 $\pm$ 1.43  \\

Learned adaptive policy (increment) & 23.20 $\pm$ 1.67 \\

Equidistant policy & 23.24 $\pm$ 1.25  \\

\label{tb:CombinedTest_d5}
\end{tabular}
\end{table}

\begin{table}[!t]
\caption{Policy Performance on Complex Combined Phantoms: A comparison of policies trained on simple phantoms, evaluated on complex combined phantoms (d6) using PSNR values.}
\centering
\begin{tabular}{c|c}

\textbf{Policies} & \textbf{Combined phantoms d6)}\\ 
\hline
Learned adaptive policy (end-to-end) & 22.35 $\pm$ 1.12  \\

Learned adaptive policy (increment) & \textbf{22.49 $\pm$ 1.27} \\

Equidistant policy & 22.20 $\pm$ 0.54  \\

\label{tb:CombinedTest_d6}
\end{tabular}
\end{table}

\begin{table*}[!t]
\caption{Performance comparison of policies on test for phantoms with different intensity values regarding the PSNR values}
\centering
\begin{tabular}{c|c|c|c|c}

\textbf{Policies} & \textbf{[0.9,1)} &   \textbf{[0.7,1)}  &  \textbf{[0.5,1)} &  \textbf{[0.1,1)}\\ 
\hline
Learned adaptive policy (end-to-end) &  \bfseries 23.91 $\pm$ 1.86 &  \bfseries 23.30 $\pm$ 2.35 & \bfseries 22.51 $\pm$ 2.91 & 21.21 $\pm$ 3.04 \\

Learned adaptive policy (increment) & 22.41 $\pm$ 2.98 &  21.41 $\pm$ 3.73  & 18.56 $\pm$ 3.92 & 17.29 $\pm$ 3.42\\

Equidistant policy &  21.77 $\pm$ 0.77 &  21.43 $\pm$ 0.76 &  21.36 $\pm$ 0.74 & \bfseries 21.31 $\pm$ 0.71 \\

\label{tb:IntensityTest}
\end{tabular}
\end{table*}

\subsection{Experiment 6 - Complex phantoms with implicit informative angles}
This study seeks to evaluate the effectiveness of the Actor-Critic agent when applied to complex, multi-material phantoms that more closely resemble realistic scenarios. The agent is trained using dataset d7) to examine its capability in a-posteriori adaptation and on dataset d8) to evaluate the impact of static components on a-priori angle selection.

\begin{figure}[t]
  \centering
  \includegraphics[width=\linewidth]{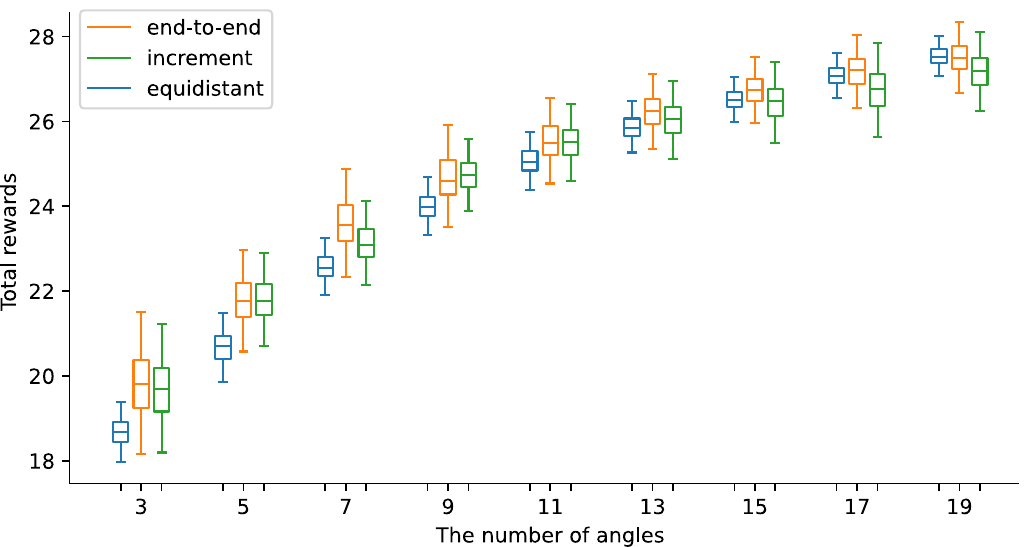}
  \caption{Comparison of policies considering different numbers of angles for the complex dataset d7): the results demonstrate the training outcomes over the last 2,000 episodes. The box represents the interquartile range in these plots, spanning from the first to the third quartile of the data distribution. The median value is displayed as a line within the box. The whiskers extend from the box to illustrate the range of the data distribution beyond the interquartile range.}
  \label{fig:SheppTraining}
\end{figure}

\begin{figure}[t]
  \centering
  \includegraphics[width=\linewidth]{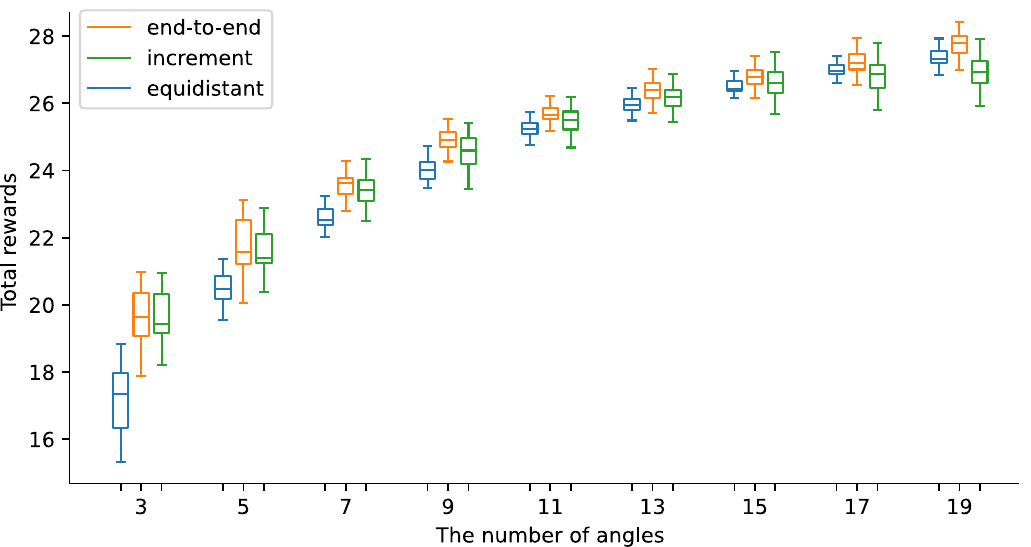}
  \caption{Comparison of policies considering different numbers of angles for the complex dataset d8): the results demonstrate the training outcomes over the last 2,000 episodes. The box represents the interquartile range in these plots, spanning from the first to the third quartile of the data distribution. The median value is displayed as a line within the box. The whiskers extend from the box to illustrate the range of the data distribution beyond the interquartile range.}
  \label{fig:SheppfixTraining}
\end{figure}

The number of phantoms used for training is 3,000 and the episodes for training are 150,000. The training outcomes during the final 2,000 episodes are illustrated in Figure (\ref{fig:SheppTraining}) for dataset d7) and Figure (\ref{fig:SheppfixTraining}) for dataset d8). Both figures affirm the superior efficacy of the Actor-Critic policies. As the number of angles increases, the differences in results among the three policies become less pronounced. This convergence suggests that with a sufficient number of angles, even an equidistant policy can achieve high-quality reconstructions. Noteworthily, the Actor-Critic policy under the end-to-end reward configuration still manifests optimal performance. Test outcomes on 300 phantoms for the unseen rotations are shown in Appendix \ref{appendix:D}.

\begin{figure}[t]
  \centering
  \includegraphics[width=\linewidth]{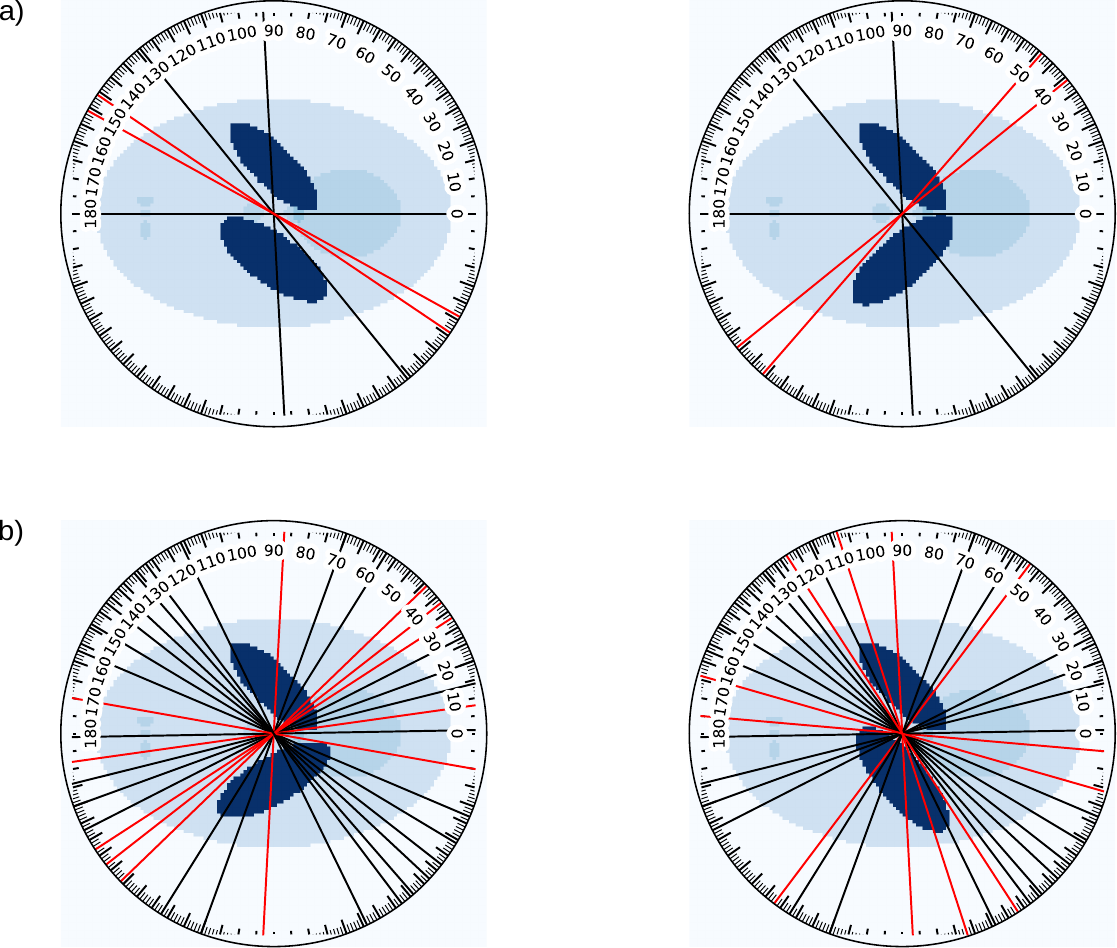}
  \caption{Illustration of personalized strategies for complex phantoms in d8) achieved using the Actor-Critic policy under the end-to-end reward setting. Two groups are depicted based on the number of angles: a) five angles; b) nineteen angles. Rows one and two present a comparison of distinct phantoms, with red lines emphasizing the a-posteriori adaptation due to ellipse rotations.}
  \label{fig:SheppFixSamples}
\end{figure}

\begin{figure}[t]
  \centering
  \includegraphics[width=\linewidth]{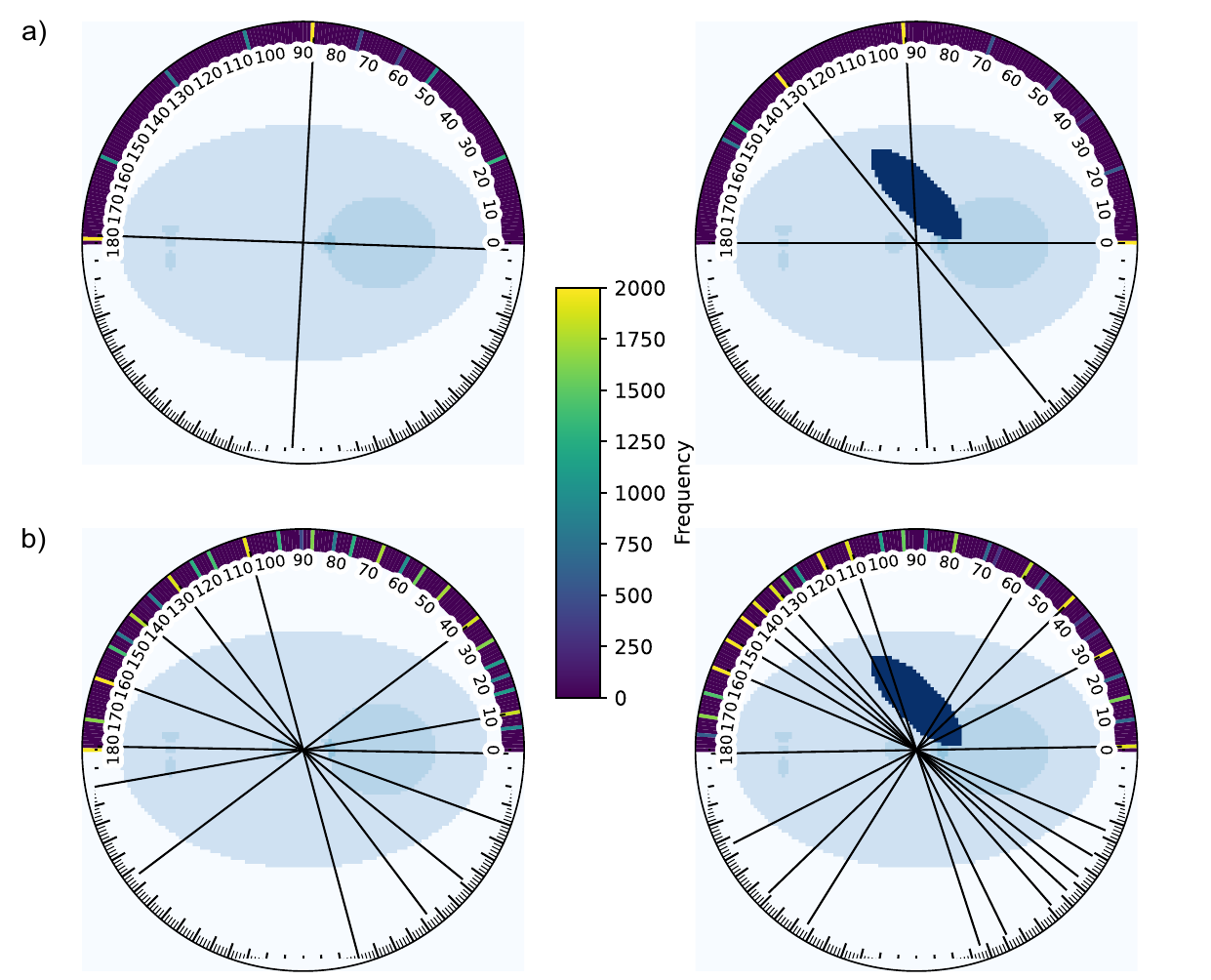}
  \caption{The figure illustrates the distribution of angles most commonly chosen, emphasizing intrinsic preferences devoid of dynamic interferences. The depicted phantoms signify the omission of any dynamic element. The gradation in color is indicative of the frequency with which specific angles were chosen over the last 2000 episodes, with each angle correlating to a color on the provided scale. Notably, angles that fall within the top frequency bracket (1750 to 2000 occurrences) are accentuated by pronounced black lines.}
  \label{fig:SheppPrior}
\end{figure}

In Figures (\ref{fig:SheppSamples}) and (\ref{fig:SheppFixSamples}), we showcase the training results for two different configurations: five angles in group a) and nineteen angles in group b). The distinction in angle choices for these configurations is emphasized by the red lines in both groups a) and b), illustrating the Actor-Critic's proficiency in a-posteriori adaptation. These red lines delineate the Actor-Critic agent's ability to monitor the rotations of the ellipses. To understand the impact of the static component on a-priori angle-selection, we juxtapose the outcomes from datasets d7) and d8) as seen in Figure (\ref{fig:SheppPrior}). In our analysis of the final 2000 episodes, angles selected over 1750 times are classified as belonging to the most frequently selected category. In comparison to group a), group b) incorporates an increased count of angles, characterized by a denser distribution, attributable to the introduction of an additional static ellipse. The reconstruction outcomes for these respective samples are detailed in Appendix \ref{appendix:C}. Collectively, these insights emphasize the Actor-Critic agent's proficiency in managing both a-priori angle selection and a-posteriori adjustments.

\begin{figure}[t]
  \centering
  \includegraphics[width=\linewidth]{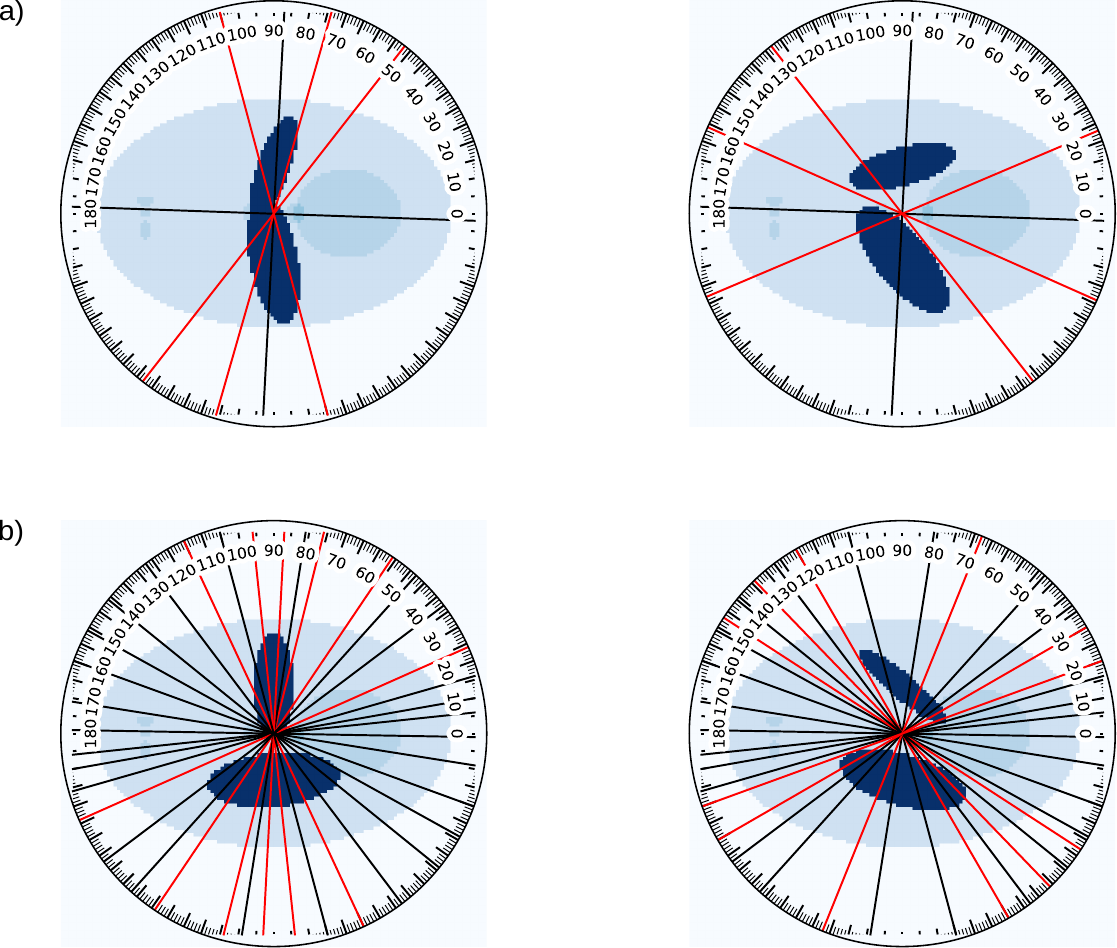}
  \caption{Illustration of personalized strategies for complex phantoms in d7) achieved using the Actor-Critic policy under the end-to-end reward setting. Two groups are depicted based on the number of angles: a) five angles; b) nineteen angles. Rows one and two present a comparison of distinct phantoms, with red lines emphasizing the a-posteriori adaptation due to ellipse rotations.}
  \label{fig:SheppSamples}
\end{figure}

\section{Discussion}
The results indicate that for both categories of phantoms—those with clear informative angles and those without—the reinforcement learning policy are able to achieve superior performance compared to the uninformed, equidistant policy. Empirical evidence suggests that the reinforcement learning policy can devise a strategy for the static component based on a-priori knowledge while authentically implementing a-posteriori adaptations. This complements the findings from \cite{shen2022learning}, whose numerical studies could not answer this important question. Furthermore, our approach simplifies the extension to other design parameters beyond angle selection. While angle selection involves straightforward sampling from one dimension of the complete dataset, altering other design parameters, such as zooming into a region of interest, introduces more complex changes in data geometry, like resolution adjustments. Designing a network to operate effectively within such a variable data environment presents significant challenges. However, in our approach, the reconstruction method intuitively accommodates these variations and seamlessly converts them to a consistent image space format, thus alleviating the complexities inherent in handling variable data geometries.

Importantly, the trained reinforcement learning policies exhibit generalization capabilities on the test dataset, including rotations not encountered during training. Introducing varying levels of noise in measurements diminishes the potential performance improvements across all policies. Yet, reinforcement learning policies continue to exhibit superior performance. This suggests that the primary challenge arises from the reconstruction of noisy data. Although the method's generalizability is somewhat limited when faced with diverse shapes and intensity values, it remains relevant under particular scenarios, especially when intensity values mirror those encountered during training. Additionally, initially trained using simpler phantoms, these policies are later evaluated on more complex ones, essentially aggregations of the simpler phantoms. This illustrates the policies' capability to adapt from basic to more intricate geometries. We have also determined that a robust training dataset is crucial. Fortunately, high-quality datasets like CAD models are accessible for industrial applications. The insights from Experiment 5 highlight the flexibility of the reinforcement learning policy in adapting to more complex or varied situations, suggesting room for further improvement.

In addition, we conducted numerical experiments to compare end-to-end and incremental reward functions. The end-to-end reward function achieves the highest average performance on both the training and test datasets. This indicates its effectiveness in guiding the reinforcement learning agent toward optimal solutions. On the other hand, the incremental reward function demonstrates faster convergence during training. In the future, we will investigate further how to design reward functions that share both of these desirable properties. 

In the future, our work can be extended in the following ways: Firstly, instead of using SIRT as an image reconstruction method, we will use deep learning-based reconstruction methods, trained end-to-end. The integration of DL in the reconstruction process is anticipated to significantly expedite and enhance the efficiency of our image reconstruction pipeline. This adaptation is expected to render the process more efficient, particularly aligning it with the demands of real-time application scenarios. Conceptually, the current separation between image reconstruction and RL involves different a-priori assumptions about the class of images being analyzed. By developing a coherent framework that unifies these assumptions for both tasks, we expect to achieve more consistent and reliable outcomes. Secondly, our forthcoming research will involve rigorous testing of our method against datasets containing defects to validate its effectiveness. Working towards this integration by incorporating defects into our phantom studies and factoring defect detection into the cost function. Thirdly, we will focus on improving the policy network to mitigate the issue of repeated angle selections.  Fourthly, we restricted ourselves to a simple 2D parallel-beam geometry to obtain scenarios in which optimal angle selection strategies are known, and the results of trained policies can be interpreted more easily. In the future, we will extend the approach to more complex and realistic 3D geometries with additional degrees of freedom, such as tilting and zooming. Fifthly, we will continue exploring and strengthening the generalization capabilities of our approach. This enhancement is crucial for ensuring that the algorithm can adapt effectively to a wide range of imaging situations. Finally, a significant aspect of our future work will involve testing and validating our algorithm on actual CT scans. This step is fundamental to transition from theoretical and simulated environments to real-world applications, and will be a critical area of focus in our continued research.

\section{Conclusion}
Compared to classical, computationally prohibitive approaches to solve the sequential OED problem of adaptive angle selection in X-ray CT, deep reinforcement learning avoids direct gradient computation on the high-dimensional, non-convex, bi-level optimization problem. Instead, it learns non-greedy strategies to solve it for a particular class of phantoms during an offline training phase which can then be applied fast and efficiently online to scans of new phantoms. We posed the sequential OED problem as a POMDP and utilized the Actor-Critic network combining a shared encoder network to learn an optimal policy. In our numerical studies with 2D CT scenarios mimicking industrial, in-line CT inspection, we could demonstrate that our approach learns efficient, truly adaptive policies that achieve better performance in terms of reconstruction quality. We introduced two different reward function settings, namely, the end-to-end and incremental reward settings. Both settings lead to stable learning processes, consolidating reinforcement learning as a reliable and extremely promising method for sequential OED. To conclude, our work demonstrates the potential of using reinforcement learning for solving sequential OED problems in inverse problems and imaging - in particular to automate angle selection and improve CT imaging efficiency, providing a flexible and adaptive approach for various CT imaging scenarios in the Industry 4.0.

\section*{Acknowledgement}

This research was co-financed by the European Union H2020-MSCA-ITN-2020 under grant agreement no. 956172 (xCTing). We would like to express our gratitude to Chat Generative Pre-trained Transformer (ChatGPT-3.5 and ChatGPT-4) for its assistance in refining the English writing of this paper.

\bibliographystyle{IEEEtran}
\renewcommand\refname{References} 

\bibliography{mybib}
\newpage
\section{Biography Section}
\begin{IEEEbiographynophoto}{Tianyuan Wang} received a Bsc. in Automation from Central South University in China and a Msc. in Computer Engineering from RWTH Aachen University in Germany. Currently, he is working as a Ph.D. student in the Computational Imaging group of the Centrum Wiskunde $\&$ Informatics (CWI) in the Netherlands. His research is part of the xCTing network and aims to realize adaptive angle selection for in-line CT.
\end{IEEEbiographynophoto}

\begin{IEEEbiographynophoto}{Felix Lucka} is a senior researcher in the Computational Imaging group at the Centrum Wiskunde $\&$ Informatica (CWI). After obtaining a first degree in mathematics and physics in 2011, he completed a PhD in applied mathematics at WWU Münster (Germany) in 2015 followed by a postdoc at University College London until 2017. His main interests are mathematical challenges arising from biomedical imaging applications that have a classical inverse problem described by partial differential equations at their core. 
\end{IEEEbiographynophoto}

\begin{IEEEbiographynophoto}{Tristan van Leeuwen} is the group leader of the Computational Imaging group at the Centrum Wiskunde $\&$  Informatica (CWI) in the Netherlands. He received his BSc. and MSc. in Computational Science from Utrecht University. He obtained his PhD. in geophysics at Delft University in 2010. After spending some time as a postdoctoral researcher at the University of British Columbia in Vancouver, Canada and the CWI, he returned to Utrecht University in 2014 as an assistant professor at the mathematical institute. In 2021, he moved to his current position. His research interests include: inverse problems, computational imaging, tomography and numerical optimization.
\end{IEEEbiographynophoto}
\vfill

\newpage
\thispagestyle{empty} 
\null

\newpage
\appendices
\section{Reconstruction samples from Experiment 4 and 5}
\label{appendix:B}
\begin{figure}[htbp]
  \centering
  \includegraphics[width=0.90\linewidth]{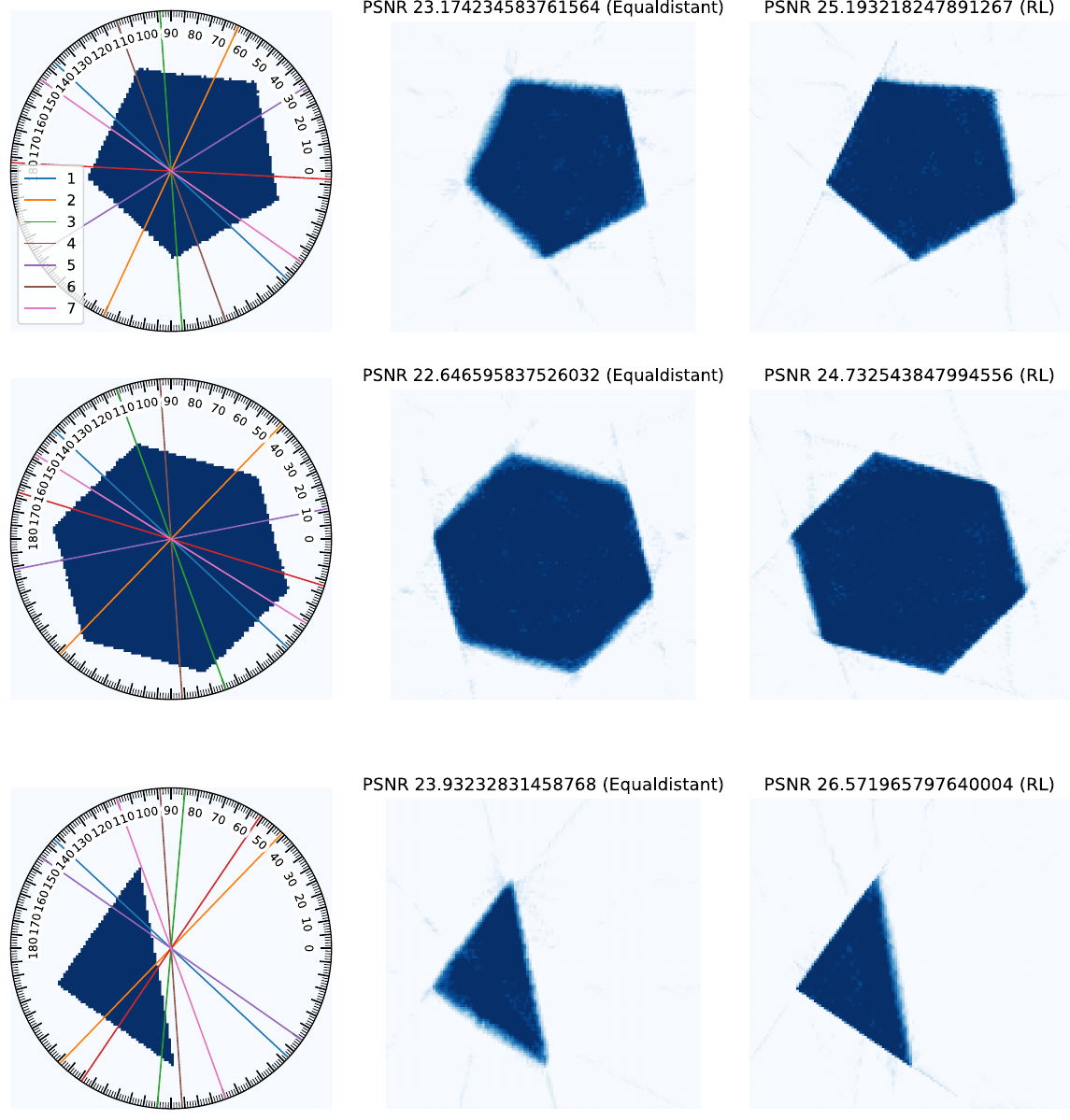}
  \caption{Theses figures illustrate reconstructions' comparison in Figure (\ref{fig:MixedSamples}). The figures on the right provide a comparative analysis of mixed phantoms. Meanwhile, the two figures on the left depict a comparison between reconstructed images obtained through the Actor-Critic policy with the end-to-end setting and those obtained through the equidistant policy.}
  \label{fig:SheppfiveRecon}
\end{figure}
\begin{figure}[htbp]
  \centering
  \includegraphics[width=0.90\linewidth]{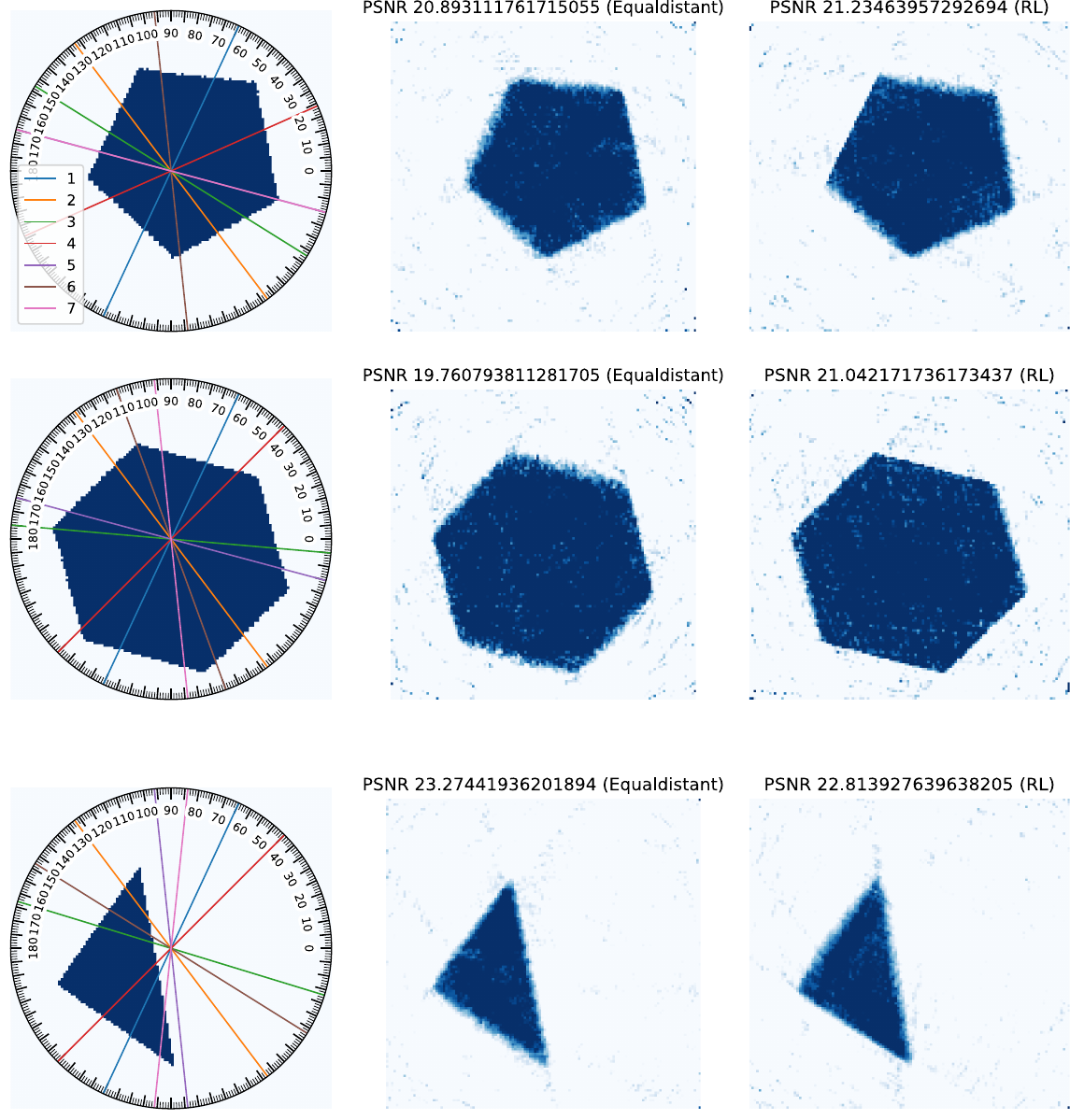}
  \caption{The figures on the right provide a comparative analysis of mixed phantoms in Figure (\ref{fig:MixedNSamples}). Meanwhile, the two figures on the left depict a comparison between reconstructed images obtained through the Actor-Critic policy with the end-to-end setting and those obtained through the equidistant policy.}
  \label{fig:SheppnineteenRecon}
\end{figure}
\FloatBarrier

\newpage

\section{Angle selection samples from Experiment 5}
\label{appendix:E}
\begin{figure}[htbp]
  \centering
  \includegraphics[width=\linewidth]{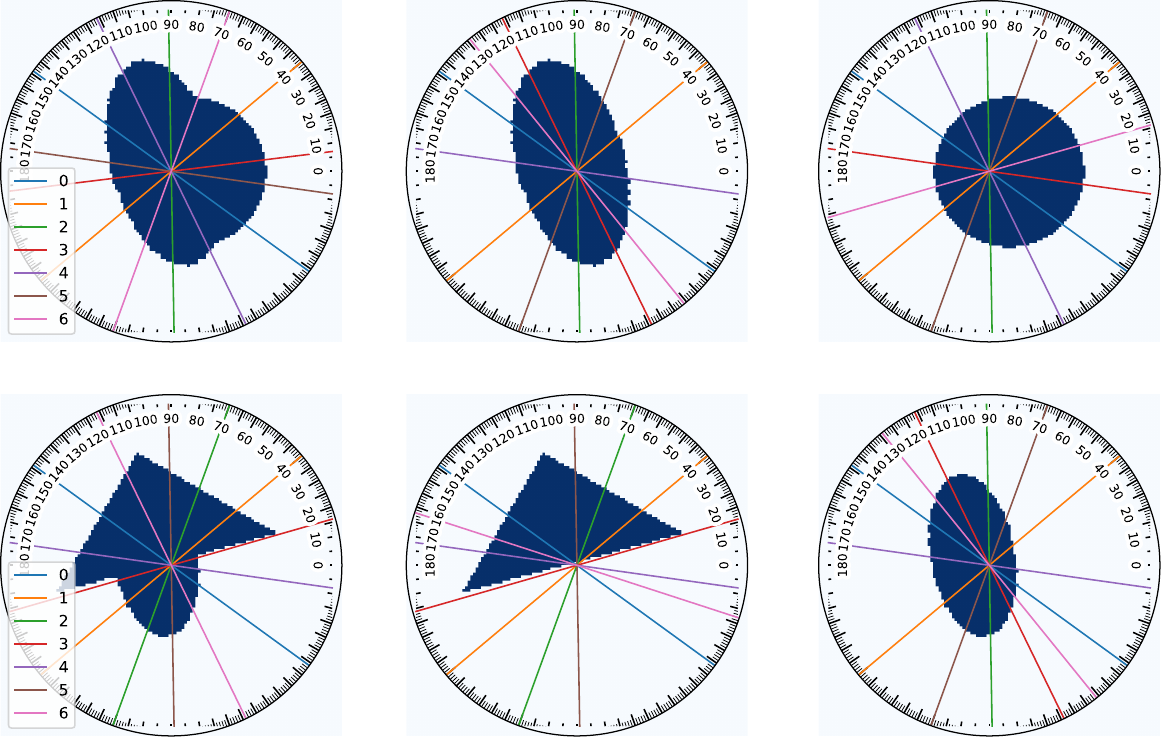}
  \caption{The figure illustrates the effectiveness of personalized strategies for two distinct groups of phantoms, as determined by the Actor-Critic policy under an end-to-end reward framework. Each row corresponds to a specific group of phantoms. For each row, the leftmost image displays the outcomes on the combined phantoms, while the center and rightmost images depict the results for individual phantoms achieved through the trained Actor-Critic policy.}
  \label{fig:GSamples}
\end{figure}

\FloatBarrier

\newpage

\section{Reconstruction samples from Experiment 6}
\label{appendix:C}
\vspace{-\baselineskip}
\begin{figure}[htbp]
  \centering
  \includegraphics[width=0.90\linewidth]{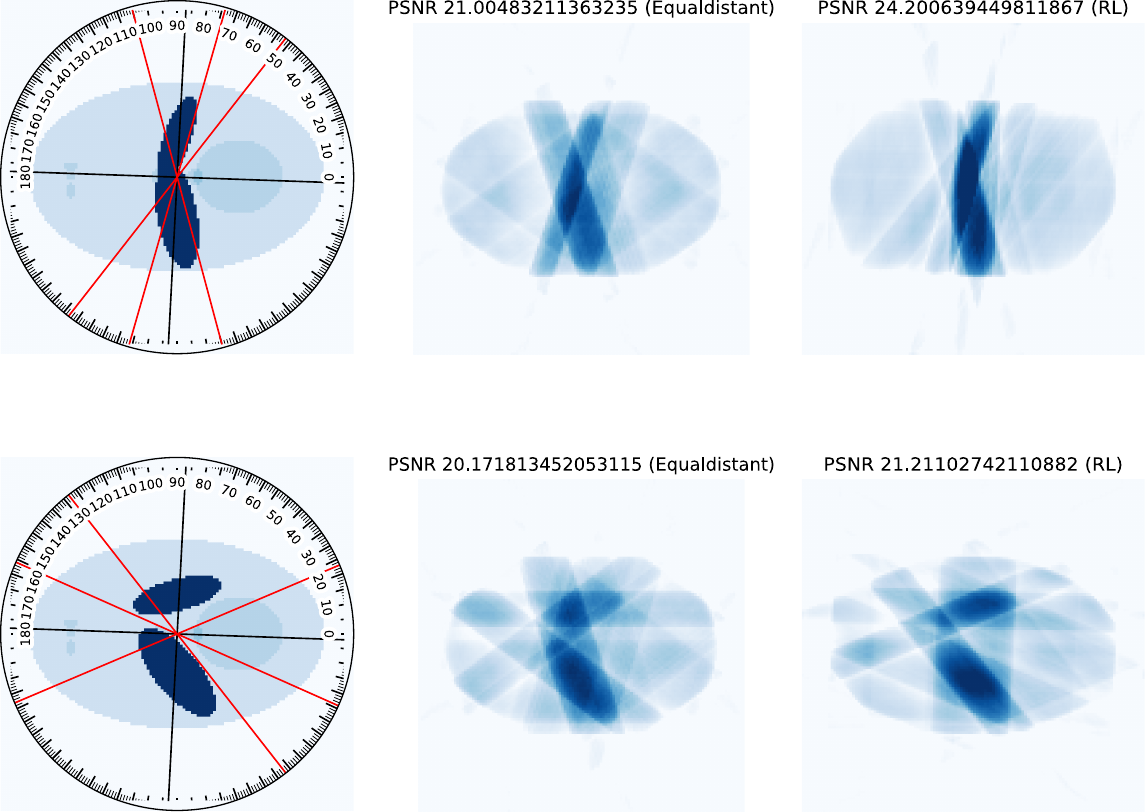}
  \caption{Theses figures illustrate reconstructions' comparison on a selection of five angles from the group a) in Figure (\ref{fig:SheppSamples}). The figures on the right provide a comparative analysis of phantoms with different positions and orientations, focusing on two ellipses. Meanwhile, the two figures on the left depict a comparison between reconstructed images obtained through the Actor-Critic policy with the end-to-end setting and those obtained through the equidistant policy. }
  \label{fig:SheppfiveRecon}
\end{figure}

\begin{figure}[htbp]
  \centering
  \includegraphics[width=0.90\linewidth]{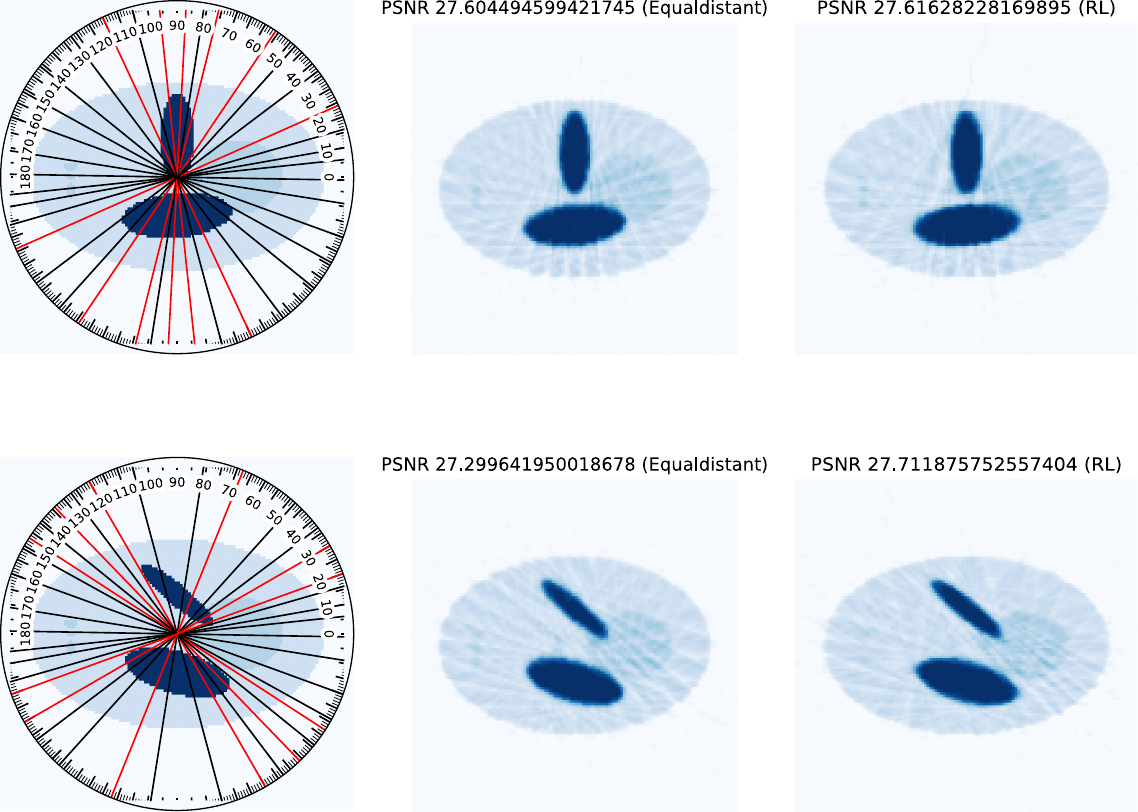}
  \caption{Theses figures illustrate reconstructions' comparison on a selection of five angles from the group a) in Figure (\ref{fig:SheppSamples}). The figures on the right provide a comparative analysis of phantoms with different positions and orientations, focusing on two ellipses. Meanwhile, the two figures on the left depict a comparison between reconstructed images obtained through the Actor-Critic policy with the end-to-end setting and those obtained through the equidistant policy.}
  \label{fig:SheppnineteenRecon}
\end{figure}

\begin{figure}[htbp]
  \centering
  \includegraphics[width=0.90\linewidth]{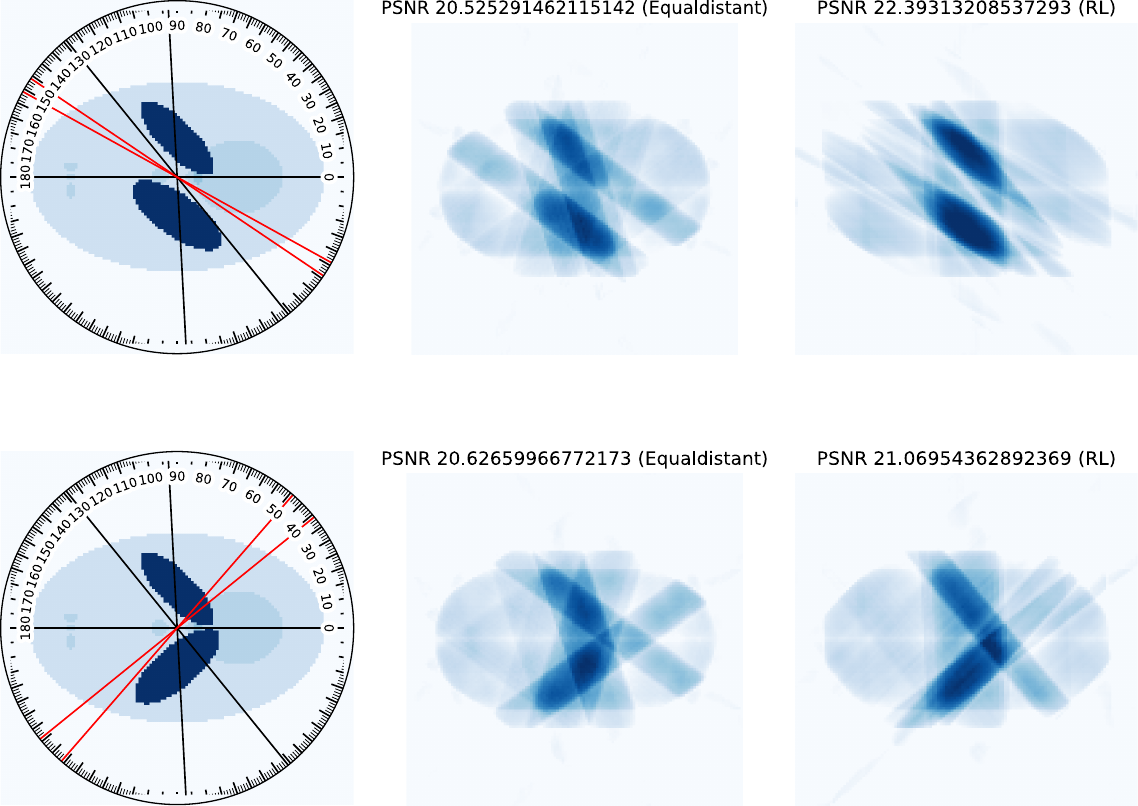}
  \caption{Theses figures illustrate reconstructions' comparison on a selection of five angles from the group a) in Figure (\ref{fig:SheppFixSamples}). The figures on the right provide a comparative analysis of phantoms with different positions and orientations, focusing on two ellipses. Meanwhile, the two figures on the left depict a comparison between reconstructed images obtained through the Actor-Critic policy with the end-to-end setting and those obtained through the equidistant policy.}
  \label{fig:SheppFixfiveRecon}
\end{figure}

\begin{figure}[htbp]
  \centering
  \includegraphics[width=0.90\linewidth]{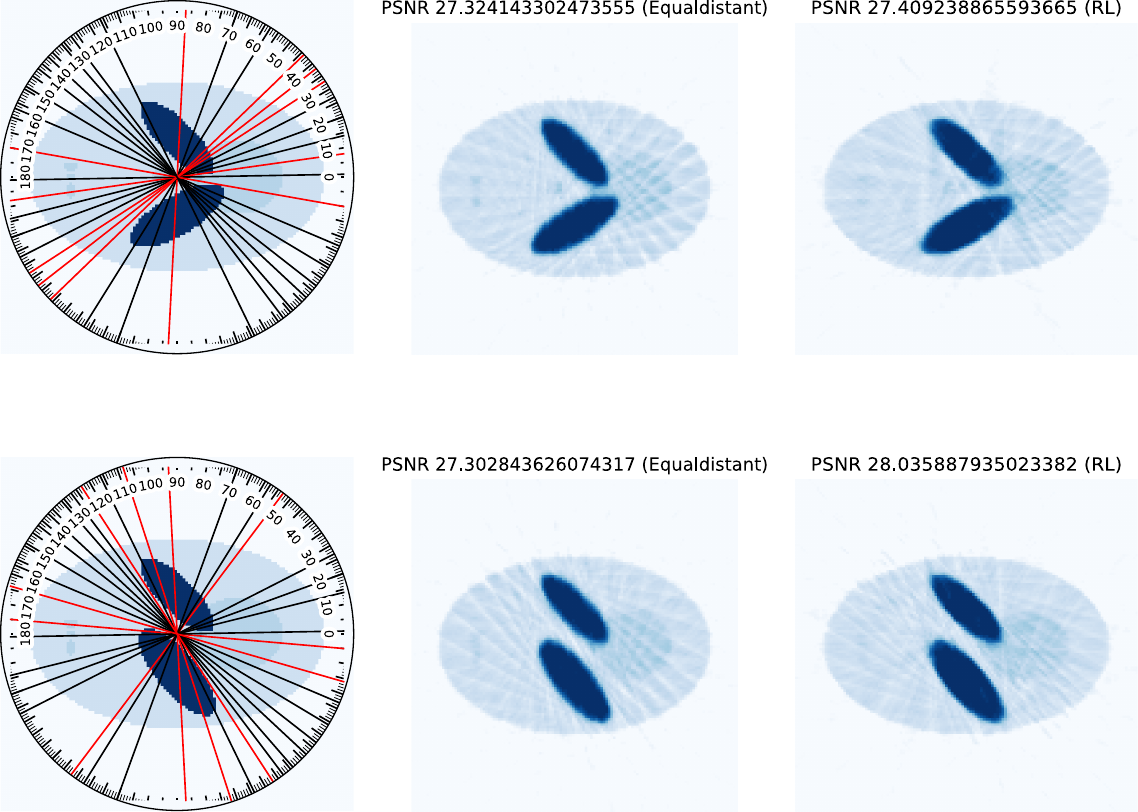}
  \caption{Theses figures illustrate reconstructions' comparison on a selection of nineteen angles from the group b) in Figure (\ref{fig:SheppFixSamples}). The figures on the right provide a comparative analysis of phantoms with different positions and orientations, focusing on two ellipses. Meanwhile, the two figures on the left depict a comparison between reconstructed images obtained through the Actor-Critic policy with the end-to-end setting and those obtained through the equidistant policy.}
  \label{fig:SheppFixnineteenRecon}
\end{figure}

\FloatBarrier
\newpage

\begin{landscape}
\section{Parameters used in datasets}
\label{appendix:A}

\begin{table}[htbp]
\caption{Parameters used in datasets}
\centering
\begin{tabular}{l|c|c}
\ & \textbf{Scales} & \textbf{Shifts}\\ 
\hline
d1) Circles & Radius: 20 $\sim$ 40 & Center coordinate: (42,42) $\sim$ (85,85)\\

d2) Ellipses & Major length: 63 $\sim$ 72, Minor length: 32 $\sim$ 40 & Center coordinate: (49,49) $\sim$ (79,79) \\

d3) Triangles & Right angled sides: 37 $\sim$ 77 &  Circumscribed circle center: (50,50) $\sim$ (70,70) \\

d4) Pentagons & Length: 44 $\sim$ 56 &  Circumscribed circle center: (54,54) $\sim$ (74,74) \\

d4) Hexagons & Length: 45 $\sim$ 50 &  Circumscribed circle center: (50,50) $\sim$ (70,70) \\

d7) Upper ellipse & Major length: 31 $\sim$ 36, Minor length: 8 $\sim$ 13 & Center coordinate: (49,49) $\sim$ (79,79) \\

 \; \quad Lower ellipse & Major length: 39 $\sim$ 41, Minor length: 15 $\sim$ 17 & - \\

d8) Lower ellipse & Major length: 39 $\sim$ 41, Minor length: 15 $\sim$ 17 & - \\

\label{tb:DataParameters}
\end{tabular}
\end{table}

\section{Unseen rotations tests from Experiment 6}
\label{appendix:D}
\begin{table}[!htp]
\caption{Performance comparison of policies on unseen rotations test for COMPLEX phantoms in d7) regarding the PSNR values}
\centering
\resizebox{\linewidth}{!}{
\begin{tabular}{c|c|c|c|c|c|c|c|c|c|c}

\textbf{Policies} & \textbf{3} & \textbf{5} & \textbf{7} & \textbf{9} &  \textbf{11} & \textbf{13} & \textbf{15} & \textbf{17} & \textbf{19}\\ 
\hline
Learned adaptive policy (end-to-end) & \bfseries 20.09 $\pm$ 0.95 & \bfseries 22.06 $\pm$ 0.85 & \bfseries 23.70 $\pm$ 0.62 & \bfseries 24.82 $\pm$ 0.71 & \bfseries 25.74 $\pm$ 0.57 & \bfseries 26.36 $\pm$ 0.43 & \bfseries 26.94 $\pm$ 0.40 & \bfseries 27.47 $\pm$ 0.47 & \bfseries 27.83 $\pm$ 0.44 \\

Learned adaptive policy (increment) & 19.97 $\pm$ 0.95 & 21.91 $\pm$ 0.61 & 23.14 $\pm$ 0.72 & 24.78 $\pm$ 0.49 & 25.57 $\pm$ 0.88 & 26.21 $\pm$ 0.66 & 26.19 $\pm$ 1.95 & 26.32 $\pm$ 2.38 & 26.59 $\pm$ 2.20 \\

Equidistant policy & 18.67 $\pm$ 0.33 & 20.69 $\pm$ 0.38 & 22.62 $\pm$ 0.32 & 24.06 $\pm$ 0.40 & 25.13 $\pm$ 0.43 & 25.95 $\pm$ 0.42 &  6.59 $\pm$ 0.40 & 27.17 $\pm$ 0.37 & 27.64 $\pm$ 0.36 \\

\label{tb:SheppTest}
\end{tabular}
}
\end{table}

\begin{table}[!htp]
\caption{Performance comparison of policies on unseen rotations test for COMPLEX phantoms in d8) regarding the PSNR values}
\centering
\resizebox{\linewidth}{!}{
\begin{tabular}{c|c|c|c|c|c|c|c|c|c|c}

\textbf{Policies} & \textbf{3} & \textbf{5} & \textbf{7} & \textbf{9} &  \textbf{11} & \textbf{13} & \textbf{15} & \textbf{17} & \textbf{19}\\ 
\hline
Learned adaptive policy (end-to-end) & 20.11 $\pm$ 0.53 & 21.89 $\pm$ 0.83 & \bfseries 23.66 $\pm$ 0.48 & \bfseries 24.83 $\pm$ 0.33 & \bfseries 25.71 $\pm$ 0.38 & \bfseries 26.52 $\pm$ 0.41 & \bfseries 27.01 $\pm$ 0.38 & \bfseries 27.59 $\pm$ 0.30 & \bfseries 28.01 $\pm$ 0.39 \\

Learned adaptive policy (increment) & \bfseries 20.11 $\pm$ 0.50 & \bfseries 21.92 $\pm$ 0.48 & 23.49 $\pm$ 0.69 & 24.64 $\pm$ 0.48 & 25.60 $\pm$ 0.43 & 26.33 $\pm$ 0.56 & 26.67 $\pm$ 0.57 & 27.10 $\pm$ 0.38 & 27.48 $\pm$ 0.41 \\

Equidistant policy & 18.93 $\pm$ 0.28 & 20.78 $\pm$ 0.28 & 22.68 $\pm$ 0.24 & 24.03 $\pm$ 0.25 & 25.24 $\pm$ 0.22 & 26.15 $\pm$ 0.24 &  26.78 $\pm$ 0.20 & 27.30 $\pm$ 0.17 & 27.66 $\pm$ 0.16 \\
\label{tb:SheppTest}
\end{tabular}
}
\end{table}

\end{landscape}

\end{document}